\documentclass[journal]{IEEEtran}

\usepackage{graphicx}
\usepackage{booktabs}
\usepackage[table]{xcolor}
\usepackage{multirow}
\usepackage{subcaption}
\usepackage{booktabs}
\usepackage{array}
\usepackage{url}
\usepackage{amsmath}
\usepackage{graphicx}
\usepackage{amssymb}
\usepackage{longtable}
\usepackage{color}
\usepackage{ragged2e}
\usepackage{tabularx}
\usepackage{xspace}

\makeatletter
\DeclareRobustCommand\onedot{\futurelet\@let@token\@onedot}
\def\@onedot{\ifx\@let@token.\else.\null\fi\xspace}
\def\eg{\emph{e.g}\onedot} 
\def\ie{\emph{i.e}\onedot} 
 
\def\etc{\emph{etc}\onedot} 
 
\def\etal{\emph{et al}\onedot}
\makeatother

\ifCLASSINFOpdf
\else
\fi
\begin{document}
\title{Interpretable and Robust AI in EEG Systems:\\
A Survey}

\author{Xinliang~Zhou,~\IEEEmembership{}
        Chenyu~Liu,~\IEEEmembership{}
        Jinan~Zhou,~\IEEEmembership{}
        Zhongruo~Wang,~\IEEEmembership{}
Liming~Zhai,~\IEEEmembership{}
Ziyu~Jia,~\IEEEmembership{}
Cuntai~Guan,~\IEEEmembership{Fellow,~IEEE}
        and~Yang~Liu,~\IEEEmembership{Senior~Member,~IEEE}% <-this % stops a space
% \thanks{Xinliang Zhou, Chenyu Liu, Liming Zhai, Cuntai Guan and Yang Liu are with the School of Computer Science and  Engineering, Nanyang Technological Univeristy, 50 Nanyang Avenue, 639798, Singapore, email: (xinliang001 and chenyu003) @e.ntu.edu.sg, (liming.zhai, ctguan and yangliu) @ntu.edu.sg }% <-this % stops a space
\thanks{Xinliang Zhou, Chenyu Liu, Liming Zhai, Cuntai Guan and Yang Liu are with the College of Computing and Data Science, Nanyang Technological Univeristy, 50 Nanyang Avenue, 639798, Singapore (email: xinliang001@e.ntu.edu.sg, chenyu003@e.ntu.edu.sg, liming.zhai@ntu.edu.sg, ctguan@ntu.edu.sg and yangliu @ntu.edu.sg).}% <-this % stops a space
\thanks{Jinan Zhou is with Carnegie Mellon University (email: jinan@cmu.edu)}
\thanks{Zhongruo Wang is with Amazon, USA (email: ysxfd@amazon.com).}
\thanks{Ziyu Jia is with the Brainnetome Center, Institute of Automation, Chinese Academy of Sciences, Beijing 100190, China and also with the University of Chinese Academy of Sciences, Beijing 100190, China (email: jia.ziyu@outlook.com).}% <-this % stops a space
% \thanks{Manuscript received April 19, 2005; revised August 26, 2015.}
}

\maketitle

\begin{abstract}
The close coupling of artificial intelligence (AI) and electroencephalography (EEG) has substantially advanced human-computer interaction (HCI) technologies in the AI era. Different from traditional EEG systems, the interpretability and robustness of AI-based EEG systems are becoming particularly crucial. The interpretability clarifies the inner working mechanisms of AI models and thus can gain user trust for transparent and ethical use of EEG systems. The robustness reflects the AI's reliability against attacks and perturbations, which is essential for sensitive and fragile EEG signals. Thus the interpretability and robustness of AI in EEG systems have attracted increasing attention, and their research has achieved great progress recently.
% However, there is still no systematic survey covering recent advances in this field, making it challenging for researchers to understand and mitigate the issues in AI-based EEG systems.
However, there is still no systematic survey of the AI-based EEG systems, making it challenging for researchers to fully grasp the state-of-the-art and emerging trends in this rapidly evolving field.
In this paper, we present the first comprehensive survey on the interpretable and robust AI techniques for EEG systems. Specifically, we first introduce the background knowledge of EEG signals. Then we propose a taxonomy of interpretability by characterizing it into three types: backpropagation, perturbation, and rule-based methods. Moreover, we categorize the robustness mechanisms into four classes: noise and artifacts, human variability, data acquisition instability, and adversarial attacks. We conduct detailed analyses and comparisons for each category. Finally, we identify several key challenges for interpretable and robust AI in EEG systems and further discuss their future directions.
This survey can provide valuable guidance for researchers to gain insights into the latest advancements and future trends in this field.
% This survey can provide valuable guidance for researchers to gain insights into the state-of-the-art and emerging trends in this rapidly evolving field.
% This survey on interpretable and robust AI in EEG systems can provide valuable guidance for researchers, enabling them to gain insights into the state-of-the-art and emerging trends in this rapidly evolving field.
\end{abstract}


% \fi
\IEEEpeerreviewmaketitle

\section{Introduction}
\IEEEPARstart{E}{lectroencephalogram} (EEG) provides valuable information about activities and states of the brain in a non-invasive way, being one of the active research areas in human-computer interaction (HCI). With the blossoming of recent artificial intelligence (AI) technologies, EEG systems have increasingly embraced the power of AI for various clinical, entertainment and social interaction applications. For example, sleep staging systems combine EEG signals with deep learning to assist physicians in rapid diagnosis \cite{eldele2021attention}. Driver monitoring systems employ EEG-based deep neural networks (DNNs) to accurately detect driver fatigue to reduce the risk of car accidents \cite{gao2019eeg}. Robotic arm control systems use DNNs to translate human thoughts (reflected by EEG signals) into control signals, helping the disabled perform basic tasks, such as drinking water or moving objects \cite{jeong2020brain}. Although significant progress has been made by AI, the AI models (especially the deep learning-based ones) still remain unexplainable due to their black-box nature and are also susceptible to intentional or unintentional attacks, raising serious concerns for the interpretability \cite{waghevaluating,ding2022tsception} and robustness \cite{nishimoto2020eeg} of AI in EEG systems.

Interpretability refers to understanding why and how the AI models make decisions and predictions. Specific to AI-based EEG systems, the interpretability allow researchers to gain insights into EEG dynamics and the link between brain states and cognitive functions, and also make it easier to identify potential biases and failure modes of EEG systems. From another point of view, the interpretability can foster user trust and acceptance of EEG systems, enabling users to build confidence in the validity and value of EEG systems.
% Otherwise, users may perceive the EEG system as a ``black box'' and be hesitant to rely on its functioning. 

Robustness refers to the degree to which the decisions and predictions of AI models are free from attacks and perturbations. Unlike traditional HCI data such as image, audio and video, EEG data derived from brain tends to be noisy and variable across individuals, resulting in a lower signal-noise ratio (SNR). This is because EEG signals are easily interfered by biological and environmental artifacts(\eg, muscle movements, eye blinks, heartbeat, electrical devices, \etc), and the same stimuli also evoke different EEG responses in different people which has unique neural rhythms.

\begin{figure}[t]
    \centering
    \begin{subfigure}{0.8\columnwidth}
        \includegraphics[width=\linewidth, trim=0cm 0cm 0cm 0cm,clip=True]{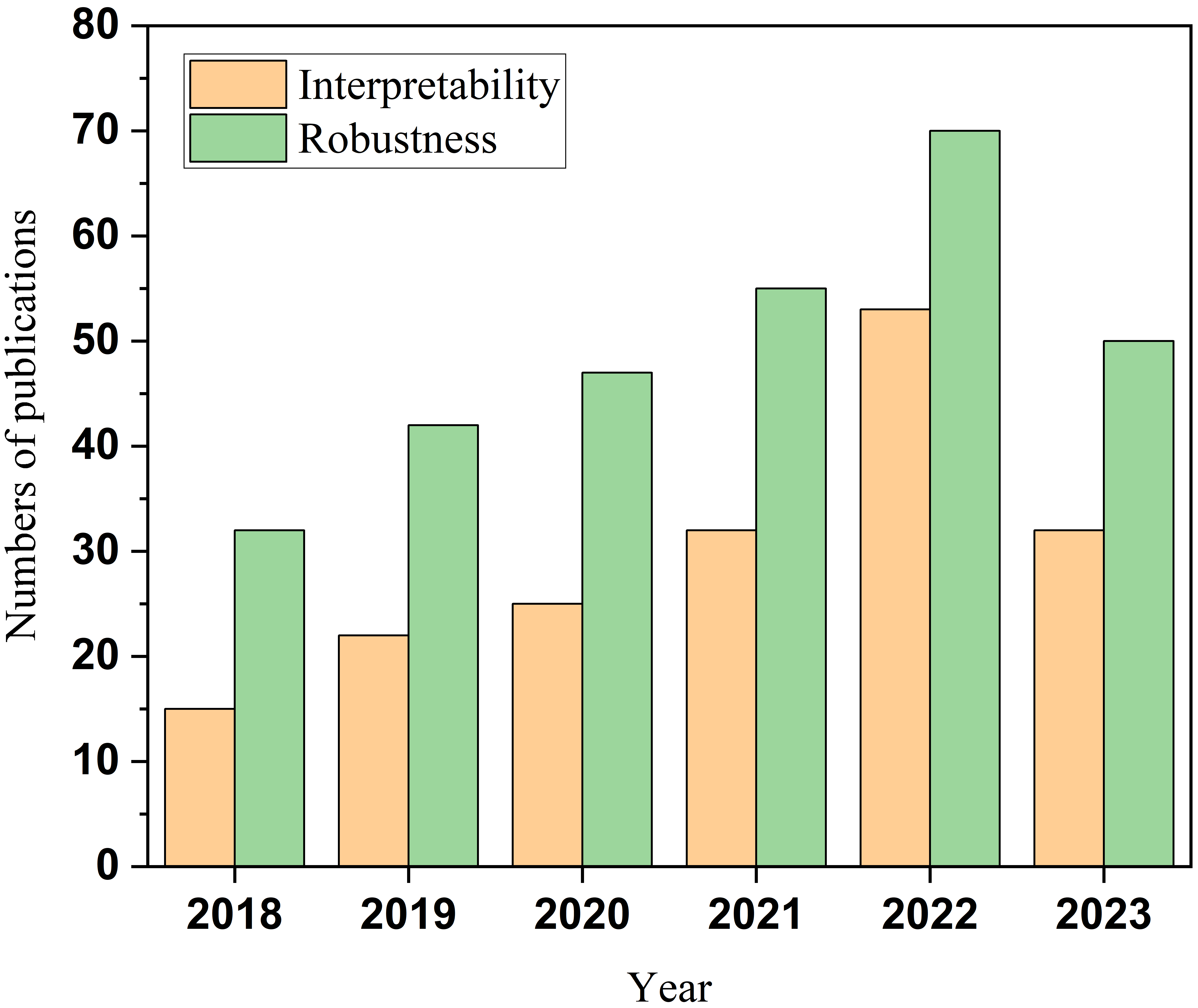}
        % \caption{Numbers of related publications}
        \caption{}
        \label{fig:numbers_of_publications}
    \end{subfigure}
    \vfill
    \vspace{5pt}
    \begin{subfigure}{0.8\columnwidth}
        \includegraphics[width=\linewidth, trim=0cm 0cm 0cm 0cm,clip=True]{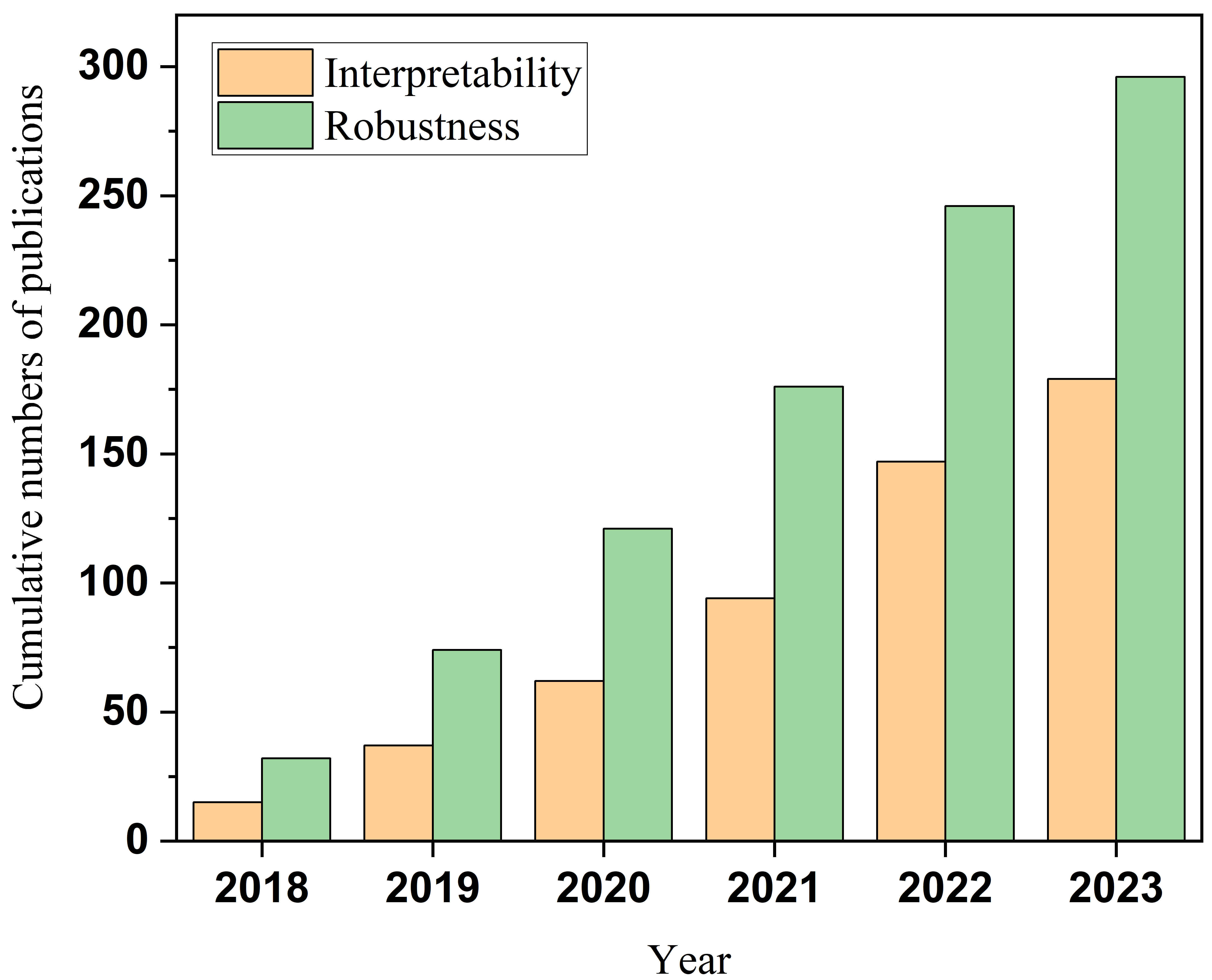}
        % \caption{Cumulative numbers of related publications}
        \caption{}
        \label{fig:cum_number}
    \end{subfigure}
    \caption{The increasing trend of interpretable and robust AI-related publications in recent six years. (a) The annual number of published papers related to interpretable and robust AI in EEG systems since 2018. (b) The cumulative number of published interpretable and robust-related papers since 2018.}
    \label{fig:numbers_publications}
\end{figure}

In recent years, there has been a surge of research interest within the academic community aimed at enhancing the interpretability and robustness of AI-based  EEG systems.
Fig. \ref{fig:numbers_publications} shows
the year-by-year number of papers on these topics.
We collect research papers employing the keywords ``interpretable'', ``interpretability'', ``explainable'', ``explainability'', ``robust'' and ``robustness'', in conjunction with ``EEG''. The selection of these papers spans from 2018 to the present and comprises publications from diverse sources such as arXiv, conference proceedings and academic journals.
The data in Fig. \ref{fig:numbers_publications} reveals that more than 60\% of these papers were published within the past two and a half years, indicating the growing popularity of this field.

While the interpretability and robustness in AI based EEG systems have raised serious concerns, and despite tremendous efforts made by researchers to address them, an exhaustive survey summarizing the state of knowledge on these two critical topics remains lacking.
There are surveys on interpretable and robust AI in general, but none of them specifically focuses on EEG Systems.
To fill this gap, in this paper we present a systematic survey covering the following aspects.

\begin{figure}[t]
\centering
\includegraphics[width=\columnwidth]{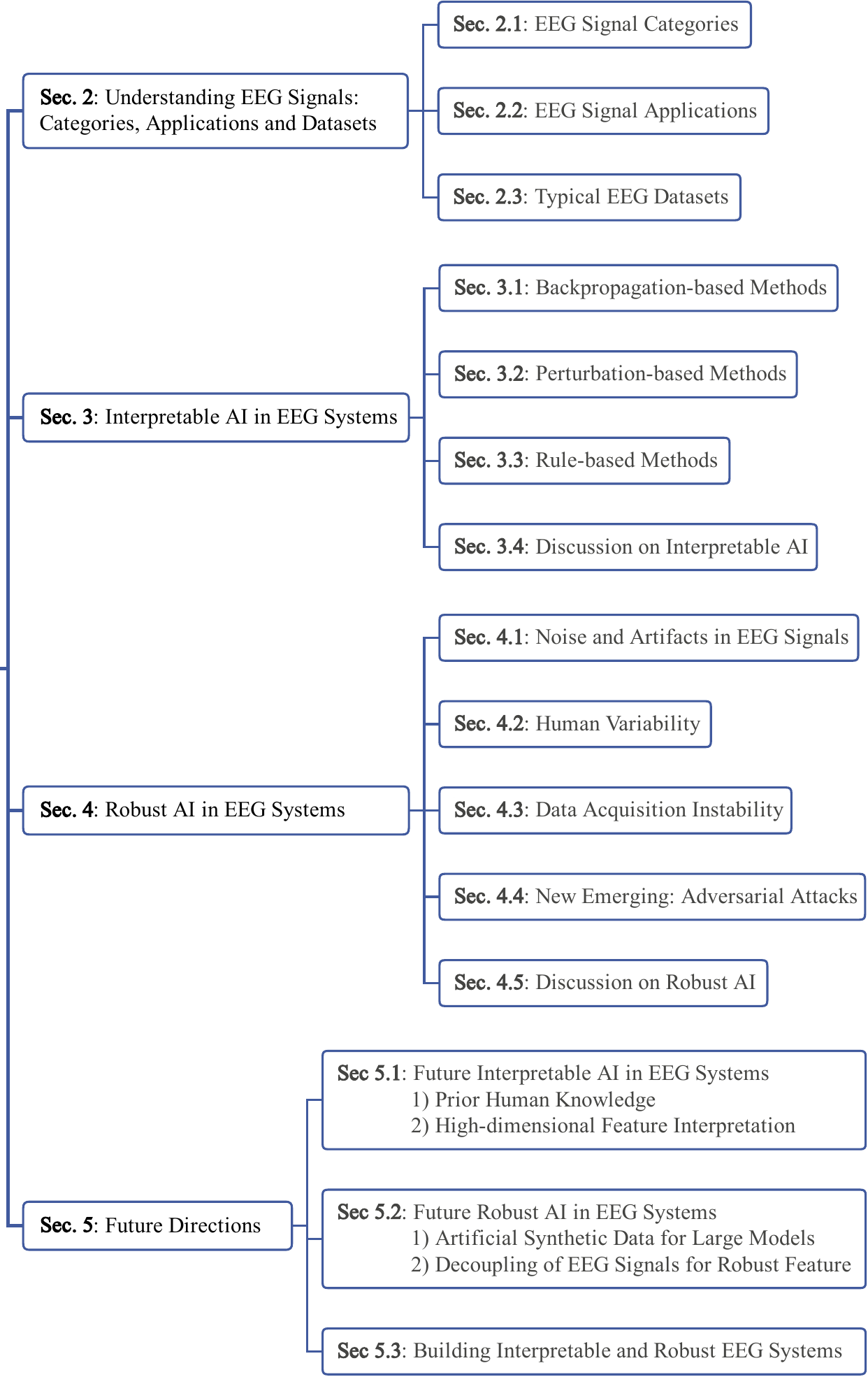}
\caption{Organization of this survey paper.}
\centering
\label{Organization}
\end{figure}

We first introduce the background of EEG signals from the EEG categories, EEG applications and EEG datasets. We then elaborate the interpretability and robustness of AI in EEG systems, respectively.
For the interpretability, we classify the interpretable AI into three types from the implementation perspective of interpretability methods. The first type is post-hoc methods based on backpropagation, which obtains the contribution of the features to results by backpropagating the prediction results. The second type is perturbation-based methods that explain initial models using local models trained with data perturbation. The last type is rule-based methods, and it applies models based on logical rules to make predictions.
For the robustness, we categorize the
robust AI into four classes: signal-component-related, subject-related, device-related, and the latest adversarial-attack-related challenges, covering all threats to the stability and security of AI-based EEG systems in practical applications.
Within each category, we summarize the common features and shared methodologies, describe representative works, and analyze their diﬀerences.
Finally, we discuss the potential directions for future research and propose practical suggestions.
The tree diagram of the paper structure is illustrated in Fig.~\ref{Organization}.

The contributions of this survey are as follows:
\begin{itemize}
  % \vspace{-10pt}
  % \setlength{\topsep}{0pt}
  % \setlength{\partopsep}{0pt}
  % \setlength{\itemsep}{-4pt}
  \item This is the first comprehensive survey focusing on the interpretability and robustness of AI in EEG systems.
  \item We propose a novel taxonomy of interpretability and robustness for EEG systems.
  \item We summarize and highlight the emerging and most representative interpretable and robust AI works related to EEG systems.
  \item We discuss some open problems and promising directions for future EEG systems. 
  % \item We provide a list of papers on the interpretability and robustness of AI in EEG systems that will be updated over time for the convenience of researchers. The list goes here: \url{https://github.com/xinliangzhou/Survey/blob/main/Paper_List.md}
  \item We maintain an ever-evolving list of papers on the interpretability and robustness of AI in EEG systems, with frequent updates to ensure its comprehensiveness and timeliness. The list goes here: \url{https://github.com/xinliangzhou/Survey/blob/main/Paper_List.md}
\end{itemize}

\section{Understanding EEG Signals:\\Categories, Applications and Datasets}
\label{sec:EEG_bkg}

In this section, we provide an overview of the EEG paradigms, including EEG signal categories, EEG signal applications and typical EEG datasets. The  EEG paradigm is a widely used approach in the field of BCI, and it involves measuring the brain's electrical activity through electrodes placed on the scalp. Compared with other paradigms in BCI, such as the ECoG paradigm \cite{ma2020comparisons}, the advantages of the EEG paradigm include its non-invasive nature, high temporal resolution, and relative ease of use. However, it has poor spatial resolution and is susceptible to interference from external sources such as muscle activity. Despite these limitations, the EEG paradigm continues to be a valuable tool in developing BCI technology.

\begin{figure}[t]
\centering
\includegraphics[width=0.8\columnwidth]{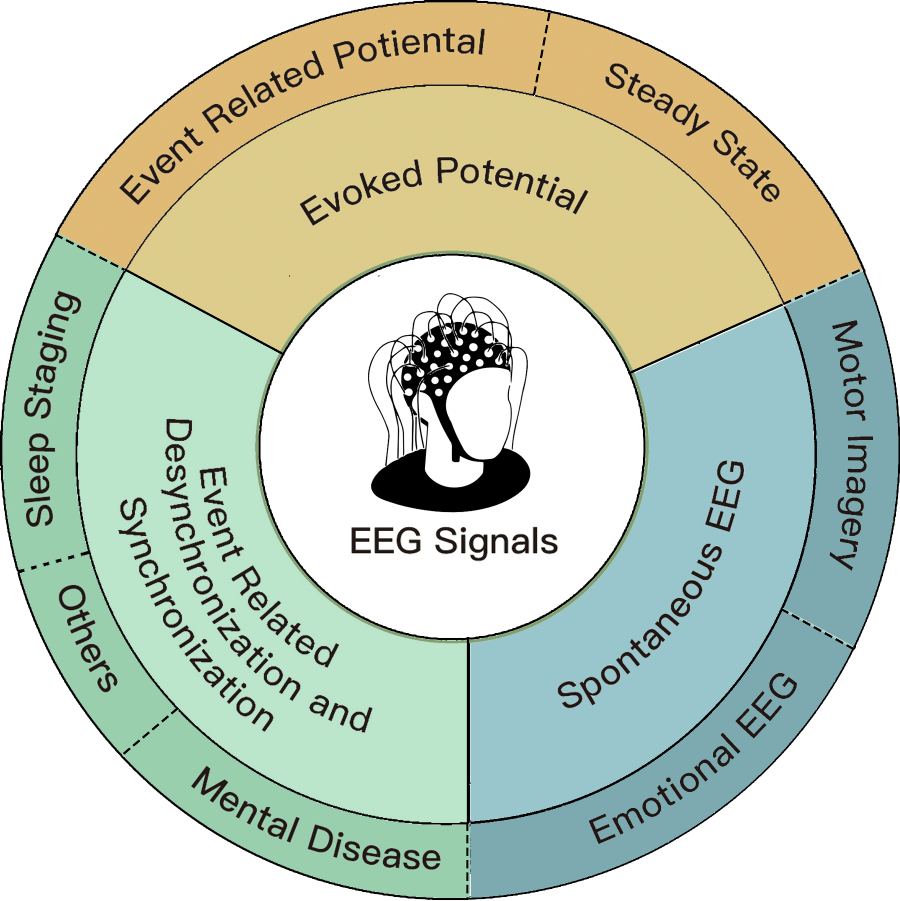}
\caption{Summary of EEG Signal Categories.}
\centering
\label{eeg_mode}
\end{figure}

\subsection{EEG Signal Categories}
% As the most frequent type of bio-signals, EEG signals contain numerous subcategories. 
% EEG signals can be divided into three categories. According to whether it is stimulated by the outside world, it is divided into spontaneous EEG and evoked potential, and a special type of event-related desynchronization/synchronization that describes mental activities.

% EEG signals can be divided into three categories. Two of them are spontaneous EEG and evoked potential, divided depending on whether subjects receive external stimuli. The other type is stimuli-irrelevant event-related desynchronization/synchronization, which describes the subject's mental activities. The EEG is classified as spontaneous EEG, evoked potential, and event-related desynchronization/synchronization, as shown in Fig.  \ref{categories}. 
%EEG signals can be divided into three categories: spontaneous EEG, evoked potentials, and event-related desynchronization/synchronization (ERD/ERS). The subjects in spontaneous EEG do not receive external stimuli, while those in evoked potentials receive external stimuli. The ERD/ERS is stimuli-irrelevant, only reflecting subjects' mental activities. The EEG signal categories are shown in Fig.  \ref{categories}.
EEG signals fall into three general categories: spontaneous EEG, evoked potentials, and event-related desynchronization/synchronization (ERD/ERS). The spontaneous EEG do not involve external stimuli presented to subjects, while the evoked potentials elicit the subjects' EEG responses to specific external stimuli. In contrast, the ERD/ERS is stimuli-irrelevant, only reflecting subjects' mental activities. The EEG signal categories are shown in Fig. \ref{eeg_mode}.

\begin{figure}[t]
\centering
\includegraphics[width=0.8\columnwidth]{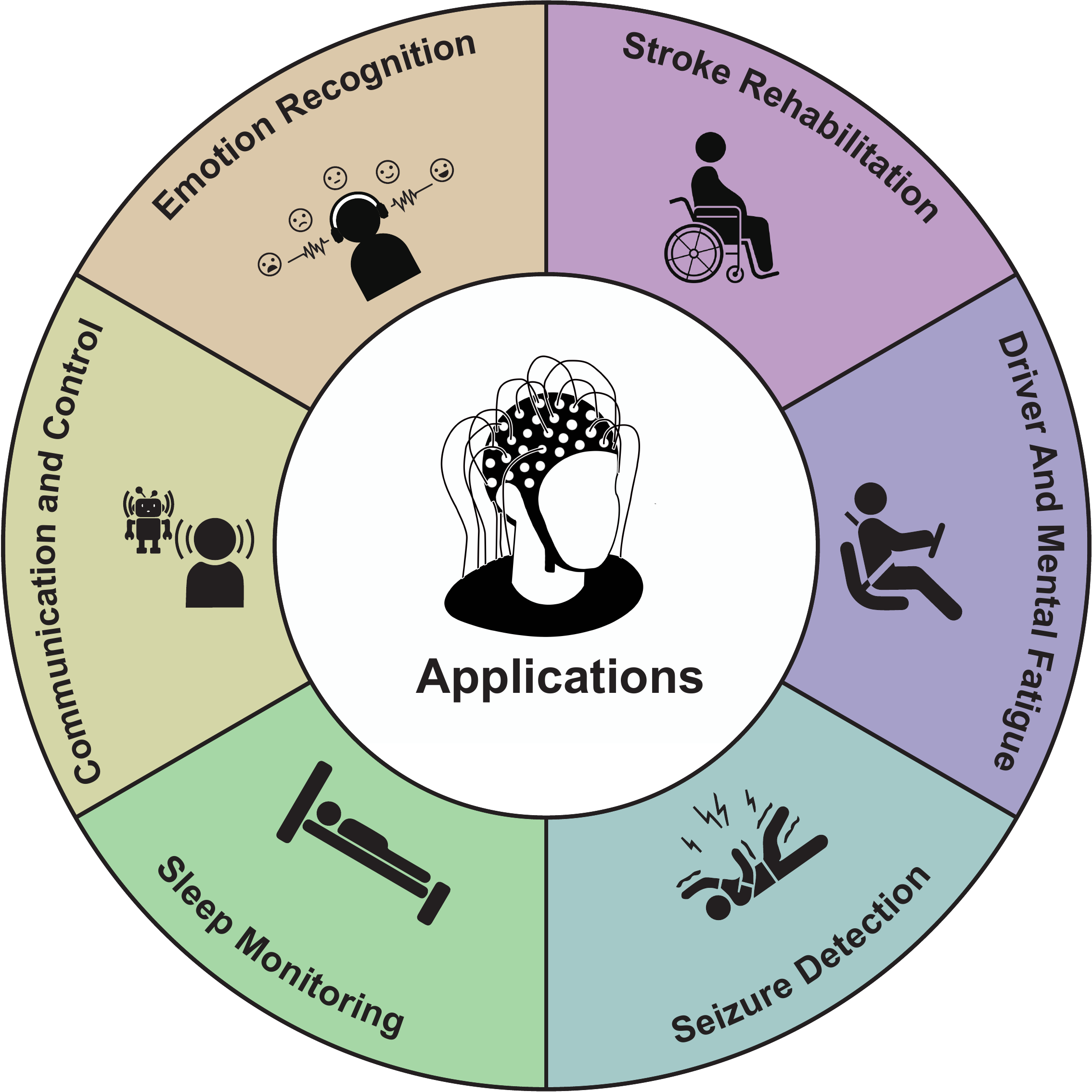}
\caption{Typical EEG Applications.}
\centering
\label{EEG_application}
\end{figure}

\subsubsection{Spontaneous EEG}
The most widely used EEG, in general, is spontaneous EEG. It refers to the measurement of brain waves obtained without external stimuli. Some common spontaneous EEG signals are obtained from the scenarios where test subjects are engaged in experiencing fatigue and sleeping, suffering from a brain disorder (\eg, Autism, Seizure), and performing motor imagery (MI) tasks \cite{hou2022gcns}. 

\subsubsection{Evoked Potentials}
Evoked Potentials (EPs), also called evoked responses, are EEG signals elicited by non-spontaneous event stimuli. Depending on different kinds of stimuli, there are two forms of EPs signals: event-related potentials (ERPs) and steady-state evoked potentials (SSEPs). The ERPs record the EEG signals elicited by specific and isolated stimulus events. While the SSEPs can reflect subjects' perception of pressure, touch, temperature and pain. On this basis, both ERPs and SSEPs contain somatosensory, auditory, and visual potentials according to the subjects' senses. All EPs signals, such as $P300$ \cite{cecotti2010convolutional}, rapid serial visual presentation (RSVP) \cite{zhang2015spatiochromatic}, and error-evoked potentials, are more robust than spontaneous signals because the amplitude and frequency of EPs are typically higher.

\subsubsection{Event-Related Desynchronization and Synchronization}  
% ERD and ERS data may be obtained using mental activities such as MI, mental arithmetic, and mental rotation. However, the performance of ERD/ERS varies widely among different users and is not remarkably accurate, making them not prevalent BCI techniques. 
The ERD/ERS reflects a relative power decrease/increase of EEG in a specific frequency band during physical motor executions and mental activities. The ERD/ERS does not require any external stimuli. However, to gather high-quality ERD/ERS signals, participants must undergo lengthy training that may last several weeks. In addition, the ERD/ERS signals are prone to fluctuate with different participants and thus have low stability.

\begin{table*}
\setlength{\extrarowheight}{0pt}
\addtolength{\extrarowheight}{\aboverulesep}
\addtolength{\extrarowheight}{\belowrulesep}
\setlength{\aboverulesep}{0pt}
\setlength{\belowrulesep}{0pt}
\caption{Summary of public datasets used in EEG systems.}
% \caption{The table offers a clear summary of diverse datasets for physiological signal processing, including emotion recognition, fatigue detection, seizure detection, sleep monitoring, and motor imagery. It lists essential attributes like dataset name, used physiological signals, participant numbers, and age ranges. This layout helps researchers understand each dataset's unique traits, aiding dataset selection for specific research. The datasets' diversity and range of signals demonstrate their robust applicability across varying demographics. Age information is included where possible.}
\label{dataset}
\centering
\resizebox{0.86\linewidth}{!}
{%
\begin{tabular}{>{\centering\hspace{0pt}}m{0.181\linewidth}>{\centering\hspace{0pt}}m{0.162\linewidth}>{\centering\hspace{0pt}}m{0.256\linewidth}>{\centering\hspace{0pt}}m{0.158\linewidth}>{\centering\arraybackslash\hspace{0pt}}m{0.181\linewidth}} 
\toprule
\rowcolor[rgb]{0.855,0.847,0.847} \textbf{Task}             & \textbf{Dataset}    & \textbf{Physiological Signal} & \textbf{Subject Number} & \textbf{Subject Age}      \\ 
\hline
\multirow{10.5}{*}{\centering\arraybackslash{}Emotion Recognition} & Deap \cite{koelstra2011deap}                & EEG, EMG, EOG, BVP             & $32$                    & $19-37$ (Mean $26.9$)                   \\
    & SEED \cite{zheng2015investigating}    & EEG    & $15$    &  Mean $ 23.27$ \\
    & SEED-IV \cite{zheng2018emotionmeter}    & EEG    & $15$    & $20 - 24$ \\
    & SEED-V \cite{liu2021comparing}    & EEG, SMI    & $20$    &  N/A \\
    & Dreamer \cite{katsigiannis2017dreamer}    & EEG, ECG    & $23$    & $20-33$ \\
    & HCI-Tag \cite{soleymani2011multimodal}    & EEG, ECG, GSR, TEMP, RSP    & $30$    & $19-40$ (Mean $26.1$) \\
    & ASCERTAIN \cite{subramanian2016ascertain}    & EEG, ECG, GSR    & $58$    & Mean $30$ \\
    & AMIGOS \cite{miranda2018amigos}    & EEG, ECG, GSR    & $40/37$      & N/A \\
    % & WESAD \cite{philip2018multimodal,greco2015cvxeda}    & ECG, EDA, EMG, RESP, TEMP    & $15$    & 28-32 \\                  
    & Enterface 06 \cite{savran2006emotion}    & EEG, fNIRS    & $16$    & N/A \\
    % & BioVid \cite{walter2013biovid}    & EEG, ECG, GSR, Pupil    & $50$    & 18-60 \\
    & Imagined Emotion \cite{onton2009high}    & EEG    & $31$    & $18-38$ \\
\hline
\multirow{3.5}{*}{\centering\arraybackslash{}Fatigue Detection}    
    % & BCI IV 2a* \cite{tangermann2012review}    & EEG, EOG    & $9$    & \emph{NA}    \\
    % & BCI IV 2b* \cite{tangermann2012review}    & EEG, EOG    & $9$    & \emph{NA}     \\
    & MEDT \cite{cao2019multi}    & EEG    & $27$    & $22-28$    \\
    & DDDE \cite{arefnezhad2022driver}    & EEG    & $13$     & $44.5 \mp 18.8$    \\
    & FatigueSet \cite{kalanadhabhatta2022fatigueset}    & EEG, ECG, PPG, EDA, TEMP     & $12$     & N/A    \\ 
\hline
\multirow{7}{*}{\centering\arraybackslash{}Seizure Detection}  
    & CHB-MIT \cite{shoeb2010application}    & EEG    & $23$     & $1.5-22$    \\
    & Bonn University \cite{andrzejak2001indications}    & EEG    & $10$     & N/A     \\
    & Freiburg Seizure \cite{ihle2012epilepsiae}   & EEG    & $21$     & Adults and Children     \\
    & Helsinki University \cite{stevenson2019dataset} & EEG    & $79$     & Infants    \\
    & EPILEPSIAE \cite{ihle2012epilepsiae}    & EEG    & $275$    & N/A    \\
    & Temple University \cite{obeid2016temple}   & EEG    & $80$     & $6-56$     \\

    % & UCI-EEG-SD    & EEG    & $22$     & 18-50    \\
    & NMT \cite{khan2022nmt}    & EEG    & N/A     & $1-90$    \\
    % & Seizure-EEG    & EEG    & $19$     & 15-65    \\

    % & CSeizure    & EEG    & $31$     & 15-55    \\ 
\hline
\multirow{8}{*}{\centering\arraybackslash{}Sleep Monitoring}     & Sleep-EDF \cite{kemp2000analysis}    & EEG, EMG, EOG    & $197$    & $18-90$    \\
    & SHHS \cite{zhang2018national}    & EEG, EMG, ECG, EOG    & $6441$    & Above $40$     \\
    & MASS \cite{o2014montreal}    & EEG, EMG, ECG, EOG    & $200$    & $18-76$    \\
    & ISRUC \cite{khalighi2016isruc}    & EEG, EMG, ECG, EOG    & $118$    & Adults and Children     \\
    & HMC \cite{alvarez2021haaglanden}    & EEG, EMG, EOG    & $105$    & $20-80$    \\
    & CAP Sleep \cite{2fbd239ff1974b7785dc3ad416a3cb0b}    & EEG, ECG, EOG    & $102$    & $18-70$    \\
    & Sleep Cassette \cite{drake1990nocturnal}      & EEG, EOG, EMG    & $100$    & $18-65$    \\
    & SIESTA \cite{penzel2011siesta}    & EEG, EOG, EMG, ECG    & $72$     & $18-80$    \\ 
\hline
\multirow{5.5}{*}{\centering\arraybackslash{}Motor Imagery}       & BCI IV 2a* \cite{tangermann2012review}    & EEG, EOG    & $9$    & N/A    \\
    & BCI IV 2b* \cite{tangermann2012review}    & EEG, EOG    & $9$    & N/A    \\
    & OpenMBI \cite{lee2019eeg}    & EEG    & $54$     & $24-45$    \\
    & Stroke \cite{ang2015facilitating, ang2014brain}    & EEG    & $21$     & Mean  $54.2$     \\
    
    & MOABB \cite{jayaram2018moabb}    & EEG    & $104$    & $18-55$    \\

\hline

\bottomrule
\end{tabular}
}
\end{table*}

\subsection{EEG Signal Applications}
% Various EEG applications, such as sleep monitoring and seizure detection, have benefited from AI technologies. The typical EEG applications based on AI technology are summarized in Fig.  \ref{EEG_application}.

% \noindent The application scenarios of the EEG system are vibrant. Here we list the five most typical applications as shown in Fig.  \ref{EEG_application}.
EEG signals have various applications. We list the six most typical applications shown in Fig.  \ref{EEG_application}.

\subsubsection{Sleep Monitoring}
Sleep monitoring plays a crucial role in the early diagnosis and intervention of sleep-related diseases. According to the sleep staging criteria proposed by the American Academy of Sleep Medicine (AASM) \cite{berry2012rules}, the complete sleep process can be divided into three stages: wakefulness (W), non-rapid eye movements (NREMs), and rapid eye movements (REMs). Different sleep stages are reflected in different prominent waveforms in EEG signals. For example, spindles and K-complex waveforms are prominent features in the NREMs stage \cite{jia2021salientsleepnet}. To save manpower and time, employing EEG signals for automatic sleep monitoring has gradually become a hot research topic \cite{jia2022hybrid,ragab2022self,liu2023bstt}.

\subsubsection{Seizure Detection}
% EEG systems are widely used in detecting and managing seizures. It can provide important information about the type and severity of a seizure and help identify the location of the seizure focus in the brain, such as the temporal lobe. This information is crucial for developing effective treatment plans and monitoring the patient's progress.
The characteristics of seizure activities can be observed from EEG signals, which provide essential information about the type and severity of the seizure and help identify the location of the seizure focus in the brain (\eg, temporal lobe) \cite{salek2006hemodynamic,bartolomei2008epileptogenicity}. The EEG signals are crucial for developing effective treatment plans and monitoring the patient's health condition.

% \begin{figure*}[h]
% \centering
% \includesvg[width=17cm]{image/AI-EEG SIGNAL UP.svg}
% \caption{EEG Signals categories}
% \centering
% \label{EEGsignal}
% \end{figure*}

\subsubsection{Fatigue Detection}
% EEG systems can be used for detecting and monitoring fatigue, such as driver fatigue and mental load fatigue. One common way is through spectral power analysis. The changes in the power of different frequency bands can be used as a useful machine-learning fatigue marker.
% Another way is through the analysis of ERP. When fatigue happens, the amplitude of certain ERP, such as the $P300$, will be reduced.
% EEG can also be used in combination with other techniques, such as eye tracking and reaction time tests, to provide a more complete picture of the cognitive and neural changes associated with fatigue.
EEG signals can be used for detecting and monitoring fatigue, such as driver fatigue \cite{cui2022eeg,liu2015brain} and mental load fatigue. One common way is through spectral power analysis, for which the changes in the power of different frequency bands can be used as fatigue indicators.
Another way is to analyze the ERPs. When fatigue happens, the amplitude of certain ERPs (\eg, $P300$) will decrease \cite{ma2014novel}. EEG signals can also be combined with other techniques, such as eye tracking and reaction time tests, providing a more complete picture of the cognitive and neural changes associated with fatigue.

\subsubsection{Communication and Control}
% Patients who have lost the majority of their motor skills (\eg, speech) are able to communicate with the outside world thanks to the aid of the EEG systems. A typical paradigm is the $P300$ speller, which enables individuals to type without any motor system and converts the user's intention into text \cite{cecotti2010convolutional}.

%  More intelligent environment stuff can be linked to EEG systems and controlled by people's brains  with the benefit of the Internet of Things (IoT). Specifically, assistive robots in smart homes, in which brain signals from individuals can control robots. 

%  Future research on brain-controlled devices may aid in developing smart homes and intelligent hospitals for the elderly and disabled.
% EEG system can decode EEG into human intent, which can be used to communicate with people or equipment. A typical paradigm is the $P300$ speller, which enables users to type without any motor system and converts their intention into text \cite{cecotti2010convolutional}. More intelligent environment stuff can be linked to EEG systems and controlled by EEG, \eg, assistive robots in smart homes. Currently, the most widely studied in this scenario is the control of robots \cite{liu2018motor}, wheelchairs \cite{yu2018asynchronous}, etc., through motor imagery. Future research on brain-controlled devices may aid in developing smart homes and intelligent hospitals for the elderly and disabled.

EEG signals representing human intent can be decoded into language or control signals used to communicate with people or intelligent devices. A typical application is the $P300$ speller, which enables users to type without any motor systems and converts their intention into text \cite{cecotti2010convolutional, thulasidas2006robust,cecotti2014single}. Besides, some intelligent environment stuff can be linked to and controlled by EEG systems, \eg, assistive robots in smart homes. These applications can be achieved by detecting SSVEP owing to the advantages of less training time, excellent recognition performance, and high information translation rate (ITR) \cite{jin2021robust,autthasan2021min2net}. Current research in this scenario focuses mainly on controlling robots \cite{liu2018motor,zhang2021improving,wang2021mvmd}, wheelchairs \cite{yu2018asynchronous,zhang2021improving}, \etc
% Future research on brain-controlled devices may aid in developing smart homes and intelligent hospitals for the elderly and disabled.

\subsubsection{Emotional Recognition}
EEG systems can be applied to assess and understand changes in the brain related to mental and physical states \cite{li2023brain,khare2020time,li2023effective}. Studies have shown differences in the activity of specific brain regions, such as the amygdala and prefrontal cortex, when a person is experiencing fear or happiness. Activities in these brain regions are directly reflected in the related EEG signals, which could be applied to mental health \cite{ha2015wearable}, cognitive psychology \cite{klimesch1996memory}, and affective computing research \cite{ding2023lggnet,ding2022tsception}.

\subsubsection{Stroke Rehabilitation}
Stroke has a high mortality rate and leads to long-term disability in up to $50\%$ of survivors. Therefore, motor rehabilitation is a top priority for post-stroke treatment \cite{arvaneh2013optimizing}. Unlike traditional stroke rehabilitation treatments, EEG systems do not rely on patients' residual motor ability. In contrast, EEG systems create a direct communication pathway between the brain and an external device \cite{zhang2022learning}, bypassing the traditional neuromuscular pathway. During EEG system-assisted rehabilitation, the system collects the patient's EEG signals and then decodes the patients' motor intentions into commands through signal processing. These commands drive the robotic device to move the patient's paralyzed limb to complete the rehabilitation exercise. Studies have reported motor cortex activation in patients who underwent EEG Systems-based rehabilitation and statistically significant improvements in patients' motor abilities during subsequent motor assessments \cite{benzy2020motor,ang2013brain}.

\subsection{Typical EEG Datasets}

The selection and utilization of datasets is a critical foundation of physiological signal processing, greatly affecting the interpretability and robustness of derived conclusions. Table~\ref{dataset} summarizes the public EEG datasets employed for various tasks, including emotion recognition, fatigue detection, seizure detection, sleep monitoring and motor imagery. 

For example, the emotion recognition task engages datasets such as Deap \cite{koelstra2011deap}, Dreamer \cite{katsigiannis2017dreamer}, and ASCERTAIN \cite{subramanian2016ascertain}, which incorporate various physiological signals, encompassing EEG, Electromyography (EMG), Electrocardiogram (ECG), Galvanic Skin Response (GSR), Temperature (TEMP), and Respiratory (RSP). These datasets cater to various age demographics and compile a substantial number of subjects, thereby augmenting the robustness of the analytic methodologies applied. Likewise, fatigue detection and seizure detection tasks leverage diverse datasets, underscoring the resilience and adaptability of the ensuing models.

Differently, sleep monitoring and motor imagery domains utilize multiple datasets collated from a wide demographic range, reinforcing the inferred models' broad applicability. The comprehensive nature of these datasets buttresses the interpretability of the findings by providing a detailed understanding of the behavior of different physiological signals under an array of conditions.

% The broad spectrum of datasets delineated in the table underscores the extensive application of these tasks across various ages and subject populations, highlighting the inherent robustness of these approaches. Concurrently, these datasets enhance interpretability by fostering a nuanced and comprehensive understanding of the discipline, thus bolstering transparency in physiological signal processing.
The broad spectrum of datasets outlined in Table~\ref{dataset} underscores the extensive application of these tasks across various ages and subject populations, highlighting the inherent robustness of these approaches. Besides, these datasets contribute to the interpretability by fostering a nuanced and comprehensive understanding of the field, thus improving the transparency in physiological signal processing.

\begin{table*}[t]
  \renewcommand{\arraystretch}{1.1}
  \newcommand{\tabincell}[2]{\begin{tabular}{@{}#1@{}}#2\end{tabular}}
  \caption{Summary of Interpretable AI in EEG Systems.}
  \label{tab1}
  \centering
  \resizebox{0.91\textwidth}{!}{
  \begin{tabular}{c|c|c|c|c}
  \noalign{\hrule height 1.1pt}
  \rowcolor{gray!20}
  \textbf{Interpetability Categories}   & \textbf{Methods} & \textbf{Coverage}     & \textbf{Explanation Type} & \textbf{Representative Works} \\
  \hline
  \multirow{4}{*}{Backpropagation-based Methods} & LRP & Local/Global & Attribution & \cite{bang2021spatio,ellis2021gradient,nagarajan2022relevance,sturm2016interpretable,lee2022quantifying} \\
  & DeepLIFT & Local/Global & Attribution & \cite{lawhern2018eegnet,gabeff2021interpreting,tahmassebi2020interpretable,jansen2018feature,tong2022tesanet} \\
  & CAM & Local & Attribution & \cite{cui2022eeg,yildiz2021classification,song2022eeg,wang2020emotion,zhang2021eeg} \\
  & Grad-CAM & Local & Attribution & \cite{liu2023fine,li2020eeg,chu2021deep,fujiwara2022deep,chen2019use} \\
  \hline
  \multirow{2}{*}{Perturbation-based Methods} & LIME & Local & Attribution & \cite{giudice2022visual,alsuradi2021trial,xu2022dagam,khare2023explainable,jeong2022novel} \\
  & SHAP & Local & Attribution & \cite{tahmassebi2020interpretable,raab2022xai4eeg,rashed2021deep,pandey2021nonlinear,ludwig2022explainability} \\
  \hline
  \multirow{3}{*}{Rule-based Methods} &
    RF & Global & Decision Rules & \cite{wang2017detection,wang2019detection,donos2015early,bentlemsan2014random,vijayakumar2017quantifying} \\
  & FIS & Global & Fuzzy Rules & \cite{feng2016joint,jiang2020eeg,ko2019multimodal,cao2017inherent,jafarifarmand2017new} \\
  & BS & Global & Bayesian Rules & \cite{qian2016drowsiness,wu2021separation,zhou2013epileptic,zhang2015sparse,zhang2013bayesian} \\
  \noalign{\hrule height 1.1pt}
  \end{tabular}
  }
\end{table*}

\section{Interpretable AI in EEG Systems}

\begin{figure}[t]
\centering
\includegraphics[width=1.0\columnwidth]{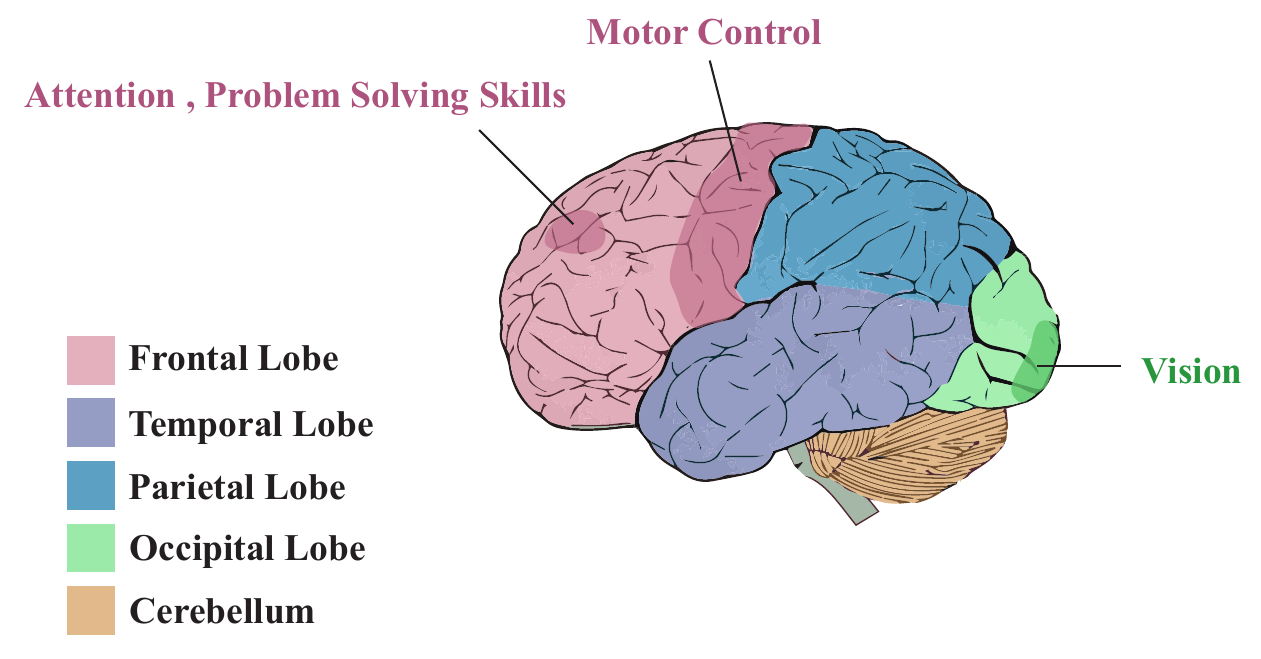}
\caption{Key brain regions related to the motor imagery task. These include the frontal lobe for cognitive skills and motor function, the temporal lobe for sensory input processing, the parietal lobe for sensory information integration, the occipital lobe for vision, and the cerebellum for coordination of voluntary movements. Each region contributes to the successful execution of the task.}
\centering
\label{MI_region}
\end{figure}

% What is Interpretable?
AI interpretability refers to explaining the decisions and actions of AI models in a manner that humans can understand.
For interpretable AI in EEG systems, it means the internal logic and workings of AI models conform to physiological principles. For example, in motor imagery (MI) tasks, the EEG signals that contribute to predictions are derived from electrodes around the motor cortex, as depicted in Fig.~\ref{MI_region}.

Interpretability is essential for EEG systems because it assesses whether the AI model has learned physiologically meaningful features. Foremost, interpretability allows checking whether the predictive logic of AI models conforms to specific proven physiological rules, since the predictive accuracy scores of the AI models can be deceptive.
For example, in MI tasks, the model making decisions may pay more attention to the noises generated by subjects' involuntary muscle movements rather than EEG signals that truly originate from cranial nerve movements.
Furthermore, interpretability methods can uncover patterns that inform brain signal research.
For example, when predicting subjects' sleep states, models identified that the signals of peripheral EEG channels generated by regular eye movements during deep sleep are highly correlated with sleep status \cite{zhu2022masking}, even though these EEG signals had long been overlooked.

After a thorough review of existing literatures, we divide the interpretability methods applied in EEG systems into three categories from the perspective of the implementation: backpropagation-based methods, perturbation-based methods and rule-based methods. We summarize the interpretability categories and their representative works in Table~\ref{tab1}.

\textbf{Type and Coverage of interpratable AI in EEG Systems:} In EEG systems, the interpretability of AI models can be broadly categorized as either local or global, influenced by feature attribution or logic rules. Local interpretability aims to explain individual predictions by illuminating why a model correlates a specific EEG pattern with a particular condition. Techniques like Layer-wise Relevance Propagation (LRP) \cite{binder2016layer}, DeepLIFT \cite{shrikumar2017learning}, Class Activation Mapping (CAM) \cite{zhou2016learning}, Gradient-weighted Class Activation Mapping (Grad-CAM) \cite{selvaraju2017grad}, Local Interpretable Model-Agnostic Explanations (LIME) \cite{ribeiro2016should}, and Shapley Additive Explanations (SHAP) \cite{lundberg2017unified} provide local interpretability.

In contrast, global interpretability illuminates the overall behavior of a model, revealing how it operates across multiple instances. Methods like random forest (RF) \cite{breiman2001random}, fuzzy inference system (FIS) \cite{aarabi2009fuzzy} and Bayesian system (BS) \cite{barber2012bayesian} are frequently utilized for the global interpretability.

% Attribution is common in methods like LRP, DeepLIFT, CAM, Grad-CAM, LIME and SHAP, assigning value to input features for a model's decision, possibly highlighting crucial brain activity patterns. Common in RF, FIS and BS, logic rules provide clear criteria for classifying EEG data conditions.
Feature attribution, assigning importance values to input features for a model's decision, is prevalent in methods such as LRP, DeepLIFT, CAM, Grad-CAM, LIME and SHAP, potentially highlighting key brain activity patterns. Meanwhile, logic rules that provide clear criteria for classifying EEG data conditions are also used in RF, FIS and BS.

% The type of explanation (attribution or logic rules) and interpretability scope (local or global) give a unique and essential insight into AI decision-making in EEG systems. The choice of method depends on the interpretability level needed and the most suitable explanation type for the data and task.
The explanation type (attribution or logic rule) and interpretability scope (local or global) provide unique and crucial insights into AI decision-making in EEG systems. The selection depends on the required level of interpretability and the most appropriate explanation type for the given data and task.

\subsection{Backpropagation-based Methods}

Backpropagation-based methods decompose the model predictions by first backpropagating the gradients from the predictions into input feature space and then visualizing the weights of these features (\eg, time-frequency patterns and electrode regions) in raw EEG signals that contribute to predictions.

\subsubsection{Layer-wise Relevance Propagation (LRP)}

% LRP provides insight into the neurophysiological phenomena behind EEG models' predictions by backpropagating prediction results. It enables EEG models to integrate information in the temporal and brain-topology-related spatial dimensions by producing LRP heatmaps.
LRP \cite{binder2016layer} provides insight into the neurophysiological phenomena behind EEG models' predictions by backpropagating results. The LRP aims to determine the contribution (measured by the relevance value) of individual elements within the input signal (corresponding to each sample point of the EEG signal) to the output prediction. It allows EEG models to integrate temporal information and brain-topography-related spatial information by producing heatmaps. 

% To implement the LRP method, a neural network, such as a convolutional neural network (CNN), must be trained initially to process EEG signals. Assuming the neural network comprises $L$ layers, each layer is assigned a weight matrix, denoted as $W^{(l)}$ with $l=1,2,\ldots,L$. The primary objective of LRP is to determine the contribution of individual elements within the input signal (corresponding to each sample point of the EEG signal) to the output prediction, termed the ``relevance'' value.
To implement the LRP, a neural network like a convolutional neural network (CNN) is first be trained to process EEG signals.
% Assuming the neural network comprises $L$ layers, each layer is assigned a weight matrix, denoted as $W^{(l)}$ with $l=1,2,\ldots,L$.
Let $R^{(l)}_i$ denotes the relevance value of neuron $i$ in layer $l$ ($l=1,2,\ldots,L$). At the output layer (\ie, $l=L$), the relevance value is equivalent to the model's predicted score:
\begin{equation}
R_i^{(L)}=f_i,
\end{equation}
where $f_i$ is the activation value of neuron $i$ in the output layer of the neural network. For each layer $l$, the backpropagation is used to propagate $R^{(l)}_i$ to the subsequent layer, $R^{(l-1)}_j$. In the LRP, this process is defined as
\begin{equation}
R_j^{(l-1)}=\sum_i \frac{f_j^{(l-1)} w_{j i}^{(l)}}{z_i^{(l)}} R_i^{(l)},
\end{equation}
where $f_j^{(l-1)}$ represents the activation value of neuron $j$ in layer $(l-1)$, $w_{ji}^{(l)}$ denotes the weight between neuron $i$ in layer $l$ and neuron $j$ in layer $(l-1)$, and $z^{(l)}_i = \sum_j f_j^{(l-1)} w_{ji}^{(l)}$ denotes the sum of the inputs to neuron $i$ in layer $l$.

% By backpropagating through each layer, the relevance values for every sample point in the input layer (\ie, the original EEG signal) can be computed. Higher relevance values suggest a more significant contribution to the prediction result, whereas lower relevance values indicate a reduced contribution. Visualization of these relevance values enables researchers to discern how the neural network extracts valuable information from EEG signals, thus offering insights for refining neural network models or interpreting brain signals in a more comprehensive manner.
Backpropagating through each layer allows us to calculate the relevance values for every sample point in the input layer (\ie, the original EEG signal). A higher relevance value indicates that the data contributed more to the prediction, while a lower value indicates less contribution. Visualizing these relevance values reveals how the neural network extracts useful information from EEG signals. This offers insights to improve neural network models or understand brain signals more thoroughly.

%%%%%%%%%%%%%%%%%%%%%%%%%%%%%%%%%%%%%%%%%%%%%%%%%%%%%%%%%%
%%%%%%%%%%%%%%%%%%%%%%%%%%%%%%%%%%%%%%%%%%%%%%%%%%%%%%%%%%

% LRP analysis ensures models concentrate on task-relevant EEG signals. Ellis \etal \cite{ellis2021gradient} apply LRP and draw heatmaps to clarify the value of local and global modalities. The results show that the distribution of local signals with higher values conforms to the regional distribution highly related to the sleep state of the human brain. Similarly, LRP could clarify the contributions of noise and neurophysiological factors. Nagarajan \etal \cite{nagarajan2022relevance} apply LRP to select the high contributing channels in MI tasks. The LRP analysis shows that the features learned by the model indeed originated from the electrodes, which are located at the action-related brain regions.
The LRP reveals whether the models focus on task-relevant EEG signals. Ellis \etal \cite{ellis2021gradient} used the LRP to generate heatmaps highlighting the local and global signals. The heatmaps show that the local signals with higher relevance values are highly related to the sleep state of the human brain, matching neurophysiological expectations. Similarly, the LRP can clarify the contributions of noise and neurophysiological factors. Nagarajan \etal \cite{nagarajan2022relevance} applied LRP to select the high contributing EEG channels in MI tasks, confirming that the model indeed learns features from the electrodes at action-related brain regions.

% LRP heatmaps explain the correlation between EEG signals from different brain regions and models' decisions at each time point. Sturm \etal \cite{sturm2016interpretable} apply LRP to track the changes of feature regions of attention of DNNs over time during action switching in MI tasks, which reveal the physiological principles behind the predictions of DNNs. What's more, both Wang \etal \cite{wang2019anxietydecoder} and Bang \etal \cite{bang2021spatio} use LRP to explain the predictions of 3D-CNN frameworks because LRP heatmaps can simultaneously reveal the contributions of frequency ranges, time intervals and spatial locations at which the signal occurs to models' predictions.
The LRP also reveals how EEG signals from different brain dimensions correlate with model decisions over time. Sturm \etal \cite{sturm2016interpretable} used the LRP to track how a DNN's attention shifted between feature regions during action switching in MI tasks, revealing the physiological principles underlying the model predictions. Moreover, Wang \etal \cite{wang2019anxietydecoder} and Bang \etal \cite{bang2021spatio} leveraged the LRP to explain 3D-CNN model predictions, for which the heatmaps simultaneously highlighted the contributions of frequency ranges, time intervals and spatial locations of relevant signals.
% Overall, LRP heatmaps uncover which EEG features influence model decisions at each moment. 

% What's more, Bang \etal \cite{bang2021spatio} use LRP to explain the predictions of 3D-CNN frameworks because LRP heatmaps can simultaneously reveal the contributions of frequency ranges, time intervals and spatial locations at which the signal occurs to models' predictions.

\subsubsection{Deep Learning Important Features (DeepLIFT)}

% Like LRP, DeepLIFT computes the correlation of each input feature with the resulting classification decision for each trial via backpropagation. DeepLIFT aims to verify whether the correct decision of the EEG model is consistent with the proven neurophysiological phenomenon and gives instructions for finding generalizable EEG features.
DeepLIFT aims to check if the model's decisions align with known neurophysiological phenomena, and provide guidance for finding generalizable EEG features. Similar to LRP, the DeepLIFT uses backpropagation to calculate how each input feature correlates with the model prediction for each trial.

% DeepLIFT is grounded in the concept of "Reference Activation," which facilitates the comparison of a feature's importance to a predefined reference point. The core principle of it is the computation of the "contribution score" for each input feature (i). The contribution score (C) can be calculated using the following equation:
The DeepLIFT is based on reference activation, enabling comparison of a feature's importance against a predefined reference point. Its core principle is to calculate a contribution score for each input feature. The contribution score can be computed using the following equation:

\begin{equation}
    C_{i} = (f_{i} - f_{i}^{0}) \times \frac{\partial y}{\partial f_{i}}, 
\end{equation}
where $C_i$ denotes the contribution score for feature $i$, $f_i$ represents the actual activation of feature $i$, $f_i^0$ signifies the reference activation for feature $i$, $\frac{\partial y}{\partial f_i}$ denotes the gradient of the output with respect to the activation of feature $i$.

For the DNNs composed of multiple layers, the chain rule is employed to compute the contribution score for each input feature. The chain rule for DeepLIFT can be expressed as
\begin{equation}
    C_{i} = \sum_{j} C_{i,j} = \sum_{j} \frac{\partial f_{j}}{\partial f_{i}}  C_{j},
\end{equation}
where $C_{i, j}$ indicates the contribution score of feature $i$ to feature $j$,
$\frac{\partial f_j}{\partial f_i}$ denotes the gradient of the activation of feature $j$ with respect to the activation of feature $i$, and $C_j$ represents the contribution score for feature $j$.

% Verifying whether the prediction logic of models conforms to physiological principles is the most common application of the DeepLIFT method in EEG systems. For example, Lawhern \etal \cite{lawhern2018eegnet} prove through DeepLIFT that their proposed EEGNet can learn to focus on EEG channels near task-related brain regions in different EEG classification tasks. Similarly, Ju \etal \cite{ju2022tensor} use DeepLIFT to interpret the spatiotemporal frequency information learned by Tensor-CSPNet in MI tasks and found it to be consistent with the fundamental frequency components existing in the left and right hands.
The most common application of the DeepLIFT method in EEG systems is to verify whether the prediction logic of models conforms to physiological principles. For example, Lawhern \etal \cite{lawhern2018eegnet} used the DeepLIFT to prove that their proposed EEGNet can learn to focus on EEG channels near task-related brain regions in different EEG classification tasks. Similarly, Ju \etal \cite{ju2022tensor} used the DeepLIFT to interpret the spatiotemporal frequency information learned by their Tensor-CSPNet in MI tasks, and found it match the key frequency components existing in the left and right hands.

% On the other hand, DeepLIFT can discover certain feature regularities from the correct predictions as a guide for brain research. For seizure detection, there is an established conclusion that high gamma frequency is an essential feature for distinguishing pre-ictal and interictal segments. However, Gabeff \etal \cite{gabeff2021interpreting} interprets the model's predictions via DeepLIFT and reveals that some low amplitude patterns are detected as ictal. This finding complements the previous conclusion that resting EEG features are also helpful as counter-balancing information in seizure detection.
On the other hand, the DeepLIFT can discover certain feature patterns from model predictions to guide brain research. For seizure detection, the high gamma frequency is known to be a key feature for distinguishing pre-ictal and inter-ictal segments. However, Gabeff \etal \cite{gabeff2021interpreting} interpreted the models using DeepLIFT, revealing that some low amplitude patterns were also detected as ictal. This finding complements the established conclusion, and verifies that the resting EEG features can also be helpful, providing counter-balancing information for seizure detection.

\subsubsection{Class Activation Mapping (CAM)}
%%%%%%%%%%%%%%%%%%%%%%%%%%%%%%%%%%%%%%%
%%%%%%%%%%%%%%%%%%%%%%%%%%%%%%%%%%%%%%%
CAM is a technique that produces heatmaps by visualizing the importance of each input feature in the final classification decision. For EEG systems, the input features are often the EEG signals from multiple channels and time points. To describe the CAM method mathematically, let $f_k(x)$ denote the activation of the $k$-th feature map in the last convolutional layer of the network, given an input $x$. The class-specific weights $w_c^k$ for class $c$ are learned during the training process. The class activation map for class $c$ can be computed as

\begin{equation}
M_c(x)=\sum_k w_c^k \cdot f_k(x),
\end{equation}
where $M_c(x)$ represents the heatmap for class $c$. This heatmap can be visualized as an overlay on the input EEG signals, highlighting the most relevant spatial and temporal features that contribute to the final classification decision.

To obtain the final classification score, the global average pooling (GAP) layer is applied on the feature maps, and then a softmax activation function is used to generate the probability distribution over classes:
\begin{equation}
p_c(x)=\frac{e^{S_c(x)}}{\sum_{c=1}^C e^{S_{c}(x)}},
\end{equation}
where $S_c(x) = \sum_{i} \sum_{j} f_k^c(x)_{i,j}$ is the sum of the activation values of the $k$-th feature map for class $c$, and $C$ denotes the total number of classes.

% CAM analysis links the high-dimensional convolutional layer features to the biologically significant features. The last convolutional neural network layer features in the work of Cui \etal \cite{cui2022eeg} are traced to bursts in the $\theta$ band and spindles in the $\alpha$ band, which strongly relate to drowsiness. Similarly, Yildiz \etal \cite{yildiz2021classification} analyze the model's prediction of the seizure by CAM and find that low-frequency EEG signals are essential in distinguishing seizures.
CAM analysis connects the deep-layered features to the biologically meaningful features. In the work of Cui \etal \cite{cui2022eeg}, the features from the last convolutional neural network layer were traced back to the bursts in the $\theta$ band and the spindles in the $\alpha$ band, which strongly relate to drowsiness. Similarly, Yildiz \etal \cite{yildiz2021classification} analyzed the seizure detection model using CAM, and found that low-frequency EEG signals are critical for distinguishing seizures.

\subsubsection{Gradient-weighted Class Activation Mapping (Grad-CAM)}

% Grad-CAM extends the Class Activation Mapping (CAM) approach by considering the gradient information flowing into the last convolutional layer of the network, offering a more precise and high-resolution visualization of relevant features in AI-based EEG systems. The Grad-CAM method computes the importance weights $\alpha_c^k$ for the $k$-th feature map in the last convolutional layer as follows:
Grad-CAM extends the CAM approach by considering the gradient information flowing into the last convolutional layer of the network, offering a more precise and high-resolution visualization of relevant features in AI-based EEG systems. The Grad-CAM method computes the importance weight $\alpha_c^k$ for the $k$-th feature map with class $c$ in the last convolutional layer as follows:

\begin{equation}
\alpha_c^k = \frac{1}{Z} \sum_{i} \sum_{j} \frac{\partial S_c(x)}{\partial f_k(x)_{i,j}},
\end{equation}
where $Z$ is the total number of spatial locations in the feature map, and $\frac{\partial S_c(x)}{\partial f_k(x){i,j}}$ denotes the gradient of the class score $S_c(x)$ with respect to the activation $f_k(x)_{i,j}$ at the spatial location $(i, j)$. The Grad-CAM heatmap for class $c$ can then be computed as

\begin{equation}
M_c^{\mathrm{Grad}}(x)=\mathrm{ReLU}\left(\sum_k \alpha_c^k \cdot f_k(x)\right),
\end{equation}
where ReLU is the rectified linear unit function, ensuring that only positive contributions are considered.
% The Grad-CAM heatmap $M_c^{\mathrm{Grad}}(x)$ can be visualized as an overlay on the input EEG signals, highlighting the most relevant spatial and temporal features contributing to the final classification decision.
Compared to the CAM, the Grad-CAM gives a more nuanced understanding of how the model behaves, since it takes into account both the positive and negative influences of the input features. 

% To gain insight into the features learned by the classifier, Fei \etal \cite{wang2020emotion} and Jonas \etal \cite{jonas2019eeg} employ Grad-CAM to reveal that higher frequency bands are particularly advantageous for emotion recognition, and identify key EEG features for EEG-based prognostication in comatose patients after cardiac arrest, respectively. Similarly, Aslan \etal \cite{aslan2022deep} employ Grad-CAM to visualize the models' outcomes and elucidate the relationship between frequency components in seizure patients and healthy individuals. On top of that, leveraging Grad-CAM for channel selection can improve decoding efficacy and strike an optimal balance between performance and channel utilization \cite{li2020eeg}. 
The Grad-CAM have been utilized to provide insight into the features learned by classifiers.
Fei \etal \cite{wang2020emotion} revealed that higher frequency bands are particularly useful for emotion recognition. Jonas \etal \cite{jonas2019eeg} identified key EEG features for prognostication in comatose patients after cardiac arrest. Similarly, Aslan \etal \cite{aslan2022deep} used the Grad-CAM to visualize model outputs and clarify the relationship between frequency components in seizure patients versus healthy individuals. Additionally, applying Grad-CAM for channel selection can enhance decoding efficacy and achieve an optimal balance between model performance and channel utilization \cite{li2020eeg}.

\subsection{Perturbation-based Methods}

% Perturbation-based methods perturb a single sample in EEG signals and observe its impact on subsequent network neurons and results to analyze the correlation between samples and model predictions. Similar to backpropagation-based methods, they are also post-hoc methods that interpret the models by attribution. The difference is that they are model-independent methods that build local models to illustrate the original models' predictive behaviors by observing local models' predictions from perturbed inputs. In other words, the local models establish the connection between biological features and initial model predictions.
Perturbation-based methods perturb individual EEG samples and observe the impact on subsequent network neurons and predictions, trying to reveal correlations between samples and model outputs. Similar to backpropagation-based methods, they are also post-hoc methods that interpret the models by attribution. However, the perturbation-based methods are model-agnostic, building local models to approximate the predictions of the original models based on perturbed inputs. In other words, the local models establish the connection between biological features and original model predictions.

\subsubsection{Interpretable Model-agnostic Explanations (LIME)}

% LIME interprets the predictions of target models by approximating them locally with naturally interpretable models. It trains local models by input perturbation and then utilizes the correlation between sample features and local model outcomes to explain the linear correlation between the physiological properties of EEG and initial model predictions.

% To be specific, LIME aims to approximate the behavior of the complex model around a specific input point $x$ by employing a simple, locally linear model, which allows for the quantification of the contribution of individual elements within the input signal to the output prediction $f(x)$. In LIME, the original input $x$ is perturbed to generate a set of similar inputs $x_i, i = 1, 2, \ldots, N$, and their corresponding predictions $f(x_i)$ are obtained from the trained neural network. Distances $d_i$ between the original input $x$ and perturbed inputs $x_i$ are computed, and weights $w_i$ are assigned to each perturbed input based on these distances, typically using an exponential kernel:

LIME explains target model predictions by approximating them locally with interpretable models. To be specific, the LIME approximates the complex model's behavior near a specific input point $x$ using a simple and locally linear model, which quantifies the contribution of individual elements within the input signal to the prediction $f(x)$. The original input x is perturbed to create similar inputs $x_i$ with $i = 1, 2, \ldots, N$, and their corresponding predictions $f(x_i)$ are obtained from the trained model. Weights $w_i$ for the perturbed inputs inputs $x_i$ are computed using an exponential kernel

\begin{equation}
w_i=\exp \left(-\frac{d_i^2}{\sigma^2}\right),
\end{equation}
where $d_i$ denotes the distances between the $x$ and $x_i$, and $\sigma$ is a scaling factor.

The simple linear model (\eg, linear regression) is then trained using the perturbed inputs $x_i$, predictions $f(x)$, and weights $w_i$. The coefficients $\beta_i$ of this simple linear model represent the contributions of each sample point in the input signal to the output prediction:

\begin{equation}
f(x) \approx \sum_{i=1}^N \beta_i x_i,
\end{equation}

% Visualizing these contributions enables researchers to gain insights into how the neural network processes EEG signals and extracts relevant information, ultimately guiding the improvement of neural network models or the interpretation of brain signals in a more comprehensive manner.
Visualizing these feature contributions provides insights into how the model processes EEG signals and extracts relevant information. This can guide researchers to improve neural network models or interpret brain signals more comprehensively. 

% Locally interpretable models provide a shortcut to directly map the predictions of initial models onto the EEG features. Giudice \etal \cite{giudice2022visual} train locally interpretable models to explain DNNs' voluntary/involuntary eye blink predictions. As demonstrated by local models, the peaks and troughs of signals, respectively, correspond to voluntary and involuntary eye blink behavior. Likewise, LIME is also employed to explain active and inactive action prediction \cite{alsuradi2021trial}. Local models display that the most significant factor in defining the action trial as active was a strong desynchronization concentrated in the $\alpha$ band and $\beta$ band. In contrast, the synchronization in these bands defines the action trial as passive. 

% Since some models contain special layers, such as SAGpooling, which result in difficulty in visualizing the contributions of different electrodes' EEG features to predictions through backpropagation. To address this problem, Xu \etal \cite{xu2022dagam} apply LIME to construct a local interpretability model to interpret the domain adversarial graph attention model (DAGAM) and find that the symmetry of EEG activities between the left and right hemispheres is a critical feature of neutral emotions.

Locally interpretable models offer a direct way to map initial model predictions onto EEG features. Giudice \etal \cite{giudice2022visual} used local models to explain DNN predictions of voluntary/involuntary blinks, revealing that the peaks and troughs in signals correspond to voluntary and involuntary eye blink behaviors, respectively. Similarly, Alsuradi \etal \cite{alsuradi2021trial} also utilized the LIME to explain active/inactive action predictions, and find that the action trial can be identified as active for strong desynchronization in the $\alpha$ and $\beta$ bands, and passive for the synchronization in those bands. 

Some models, containing specialized layers like SAGpooling, impede the visualization of feaure contributions through backpropagation. To tackle this issue, Xu \etal \cite{xu2022dagam} applied the LIME to construct local interpretability models for the domain adversarial graph attention model (DAGAM). They identify that the symmetry of EEG activities between the left and right hemispheres is a critical feature of neutral emotions.

\subsubsection{Shapley Additive Explanation Values (SHAP)}

SHAP quantifies the contribution of each input features to prediction based on the Shapley values from game theory. The Shapley value refers to the marginal contribution of the EEG feature, which is the difference between prediction results before and after the feature is added. The SHAP value for feature $i$ in a model $f$ is defined as
\begin{equation}
% \resizebox{0.90\columnwidth}{!}
% {$
\phi_i(f) \!=\! \sum_{S \subseteq N \backslash \{i\}} \frac{|S| !(|N| \!-\! |S| \!-\! 1) !}{|N| !} [f(S \cup \{i\}) \!-\! f(S)],
% $},
\end{equation}
where $N$ is the set of all input features, $S$ is a subset of features without feature $i$, and $|S|$ denotes the cardinality of $S$. The term $f(S \cup \{i\}) - f(S)$ represents the marginal contribution of feature $i$ when added to the subset of features $S$.

% SHAP values satisfy three desirable properties: local accuracy, missingness, and consistency. Local accuracy ensures that the sum of the SHAP values for each input feature and the expected model output equals the model's prediction for a specific instance. Missingness states that if a feature is missing or has no impact on the model's output, its corresponding SHAP value will be zero. Consistency guarantees that if a model assigns a higher contribution to a feature in a new model compared to an old one, the SHAP value of that feature should not decrease.

% SHAP is usually utilized to explain complex AI models in EEG systems. Tahmassebi \etal \cite{tahmassebi2020interpretable} constructs a deep neural network with the real-time predictive capability to monitor the patient's eye state. To consider models' interpretability in practical applications, the authors use SHAP to build local interpretable models to establish the relationship between EEG features in the deep neural network and the patient's state. Similarly, Raab \etal \cite{raab2022xai4eeg} also utilize SHAP to explain features' contributions of initial models with local interpretable models. The authors interpret two deep learning models of different dimensions ($1$-D and $3$-D) through SHAP for seizure detection.

SHAP values have three key properties: local accuracy, missingness, and consistency. Local accuracy ensures that the sum of SHAP values for each input feature and the expected model output equals the model prediction for a specific instance. Missingness idictates that if a feature is missing or has no impact on the model prediction, its SHAP value will be zero. Consistency guarantees that if a feature contributes more in a new model compared to an old one, the SHAP value of that feature should not decrease.

The SHAP is often utilized to explain complex AI models in EEG systems. Tahmassebi \etal \cite{tahmassebi2020interpretable} constructed a real-time DNN model to monitor patients' eye states. To ensure model interpretability in practical scenarios, they employ the SHAP to build locally interpretable models that reveal the relationship between EEG features and the eye state. Raab \etal \cite{raab2022xai4eeg} also used the SHAP to explain the feature contributions of initial models, which are DNN models of different dimensions ($1$-D and $3$-D) for seizure detection.

% \begin{figure*}[t]
% \centering
% \includegraphics[width=1.8\columnwidth]{image/interpretable_compare.pdf}
% \caption{This table provides a comparative overview of various interpretability methodologies applied in EEG systems. Key aspects such as notable benefits, suggested use cases and inherent limitations of each method are underscored. This summary serves as a vital guide for researchers, aiding in selecting the most suitable interpretability techniques for specific EEG system analysis scenarios.}
% \centering
% \label{tab:cmp_interpre}
% \end{figure*}

\begin{table*}[t]
  \renewcommand{\arraystretch}{1.1}
  \newcommand{\tabincell}[2]{\begin{tabular}{@{}#1@{}}#2\end{tabular}}
  \caption{Comparison of different interpretability methods in EEG Systems.}
  \label{tab:cmp_interpre}
  \centering
  %\resizebox{\textwidth}{!}
  {
  % \begin{tabular}{c|m{5cm}<{\centering}|m{5cm}<{\centering}|m{5cm}<{\centering}}
  \begin{tabular}{c|m{4.6cm}<{\centering}|m{4.6cm}<{\centering}|m{4.6cm}<{\centering}}
  \noalign{\hrule height 1.1pt}
  \rowcolor{gray!20}
  & \textbf{Backpropagation-based Method} & \textbf{Perturbation-based Method} & \textbf{Rule-based Method} \\
  \hline
  Mechanism & Analyze the feature contribution by backpropagating the gradients from predictions. & Explain the original model's behavior with local surrogate models. & Explain model using specific logic rules \\
  \hline
  Explanation Stage & Post-hoc & Post-hoc & Ante-hoc \\
  \hline
  Model Dependence & Model-specific & Model-agnostic & Model-agnostic \\
  \hline
  Flexibility & Low & High & High \\
  \hline
  Application Scenario & Differentiable models & Tolerable of high computational costs & Availability of priori knowledge \\
  \hline
  Limitation & Gradient dependency;
  Narrow applicability & Computationally intensive; Prone to overfitting & Oversimplify complex EEG systems; Require domain expertise \\
  \noalign{\hrule height 1.1pt}
  \end{tabular}
  }
\end{table*}

\subsection{Rule-based Methods}

Perturbation-based methods perturb individual EEG samples and observe the impact on subsequent network neurons and predictions, trying to reveal correlations between samples and model outputs. Similar to backpropagation-based methods, they are also post-hoc methods that interpret the models by attribution. However, the perturbation-based methods are model-agnostic, building local models to approximate the predictions of the original models based on perturbed inputs. In other words, the local models establish the connection between biological features and original model predictions.

\subsubsection{Random Forest (RF)}
RF is an rule-based method based on the ``IF-THEN-ELSE'' logic rule (\ie, decision rule).
The interpretability of RF arises from two main aspects: feature importance and decision paths.

\textbf{Feature Importance}: Feature importance in RF is typically calculated using the Gini importance or mean decrease impurity (MDI). For each feature $i$ derived from the EEG signals, the Gini importance $I_G(i)$ is given by

\begin{equation}
I_G(i)=\sum_{t \in T_i} \frac{n_t}{n}\left(1-\sum_{c=1}^C p_{t c}^2\right),
\end{equation}
where $T_i$ represents the set of nodes in the RF that split on feature $i$, $n_t$ is the number of samples reaching node $t$, $n$ is the total number of samples, $C$ is the number of classes, and $p_{tc}$ is the proportion of samples with class $c$ in node $t$. The Gini importance measures the decrease in node impurity, which is weighted by the probability of reaching each node.
The features of the EEG signals can be ranked based on their Gini importance, so that the most relevant features (also the most relevant EEG signals) contributing to the model's decision-making process can be identified. 
% By ranking the features based on their Gini importance, researchers can identify the most relevant features in the EEG signals that contribute to the model's decision-making process.

\textbf{Decision Paths}:
RF consists of multiple decision trees, each of which is trained on a bootstrapped sample of the original data. The decision path of an instance in a tree is the sequence of nodes from the root to a leaf node, which corresponds to the class assigned by the tree.

% By analyzing the decision paths of an instance across all the trees in the random forest, researchers can gain insights into the decision-making process and identify the common patterns or rules that lead to specific predictions.
RF model interpretability can be achieved by analyzing decision paths for an instance across all trees. This reveals common patterns and rules leading to predictions, providing insights into the decision-making process. 

% Visualizing feature importance and decision paths allows researchers to better understand the relationship between input EEG signals and the model's predictions. This understanding can help in refining the model or interpreting brain signals more comprehensively, ultimately contributing to the advancement of research in the EEG domain.

% For instance, Abdulhay \etal \cite{abdulhay2017classification} classify the shannon entropy of the instantaneous values associated with each intrinsic mode function (IMF) by RF, directly responding to the association between the instantaneous amplitudes and frequencies of the IMF and seizures. Li \etal \cite{li2021neural} also use a similar approach for seizure classification. The difference is that the authors classify the electrodes directly by RF to find the seizure onset zone (SOZ).
Visualizing feature importance and decision paths provides a deeper understanding of how input EEG signals relate to model predictions, and can aid model refinement and more comprehensive brain signal interpretation. For example, Abdulhay \etal \cite{abdulhay2017classification} employed the RF to classify the Shannon entropy of instantaneous values associated with each intrinsic mode function (IMF), directly responding to the association between the instantaneous amplitudes and frequencies of the IMF and seizures.  Li \etal \cite{li2021neural} followed a similar approach for seizure classification. The difference is that the authors classify the electrodes directly by RF to find the seizure onset zone (SOZ).

\subsubsection{Fuzzy Inference System (FIS)}
% The FIS is a computational framework based on the fuzzy set theory, fuzzy logic, or fuzzy reasoning, offering inherent interpretability due to the human-readable rules and transparent reasoning processes. In the context of AI EEG systems, it can represent the relationship between EEG features and prediction in a natural language.
FIS is a computational framework based on the fuzzy set theory, fuzzy logic or fuzzy reasoning. It offers inherent interpretability by using human-readable rules and transparent reasoning processes.

A typical FIS consists of four main components: fuzzification, fuzzy rule base, fuzzy inference engine, and defuzzification.
The \emph{fuzzification} involves converting crisp input values into fuzzy sets using membership functions. For each input variable $x_i$, a membership function $\mu_{A_i}(x_i)$ is used to determine the degree of membership of $x_i$ to the fuzzy set $A_i$:
\begin{equation}
\mu_{A_i}\left(x_i\right): x_i \rightarrow[0,1],
\end{equation}
where $\mu_{A_i}(x_i)$ represents the degree of membership of $x_i$ to the fuzzy set $A_i$.

The \emph{fuzzy rule base} is a collection of human-readable ``IF-THEN'' rules that describe the relationships between input and output fuzzy sets. A fuzzy rule can be expressed as:
\begin{equation}
\resizebox{0.85\columnwidth}{!}{$
R_k: \text { IF } x_1 \text { is } A_{k 1} \text { AND } \cdots \text { AND } x_n \text { is } A_{k n} \text { THEN } y \text { is } B_k \text {, }
$}
\end{equation}
where $x_i$ are the input variables, $A_{ki}$ and $B_k$ are fuzzy sets, and $R_k$ represents the $k$-th fuzzy rule.

The \emph{fuzzy inference engine} combines the fuzzified input values and fuzzy rules to produce fuzzy output sets. The firing strength or weight $w_k$ of each rule $R_k$ is computed as the product of the membership degrees of the input values to their corresponding fuzzy sets:
\begin{equation}
w_k=\prod_{i=1}^n \mu_{A_{k i}}\left(x_i\right),
\end{equation}

The fuzzy output sets are then generated by aggregating the weighted consequent fuzzy sets $B_k$:
\begin{equation}
B_k^{\prime}(x)=w_k \cdot B_k(x),
\end{equation}
where $B'_k(x)$ is the weighted fuzzy output set for rule $R_k$.

% The \emph{defuzzification} process converts the fuzzy output sets back into crisp output values. One common method is the centroid defuzzification, which calculates the centroid or the weighted average of the aggregated fuzzy output sets:
% \begin{equation}
% y=\frac{\sum_x B_k^{\prime}(x) \cdot x}{\sum_x B_k^{\prime}(x)},
% \end{equation}
% where $y$ is the crisp output value.
The \emph{defuzzification} process converts the fuzzy output sets back into crisp output values. One common method is the centroid defuzzification, which obtains the crisp output value $y$ by calculating the centroid of aggregated fuzzy output sets
\begin{equation}
y=\frac{\sum_x B_k^{\prime}(x) \cdot x}{\sum_x B_k^{\prime}(x)}.
\end{equation}

% The insights gained from these human-readable rules can help refine the model or interpret brain signals more effectively. Feng \etal \cite{feng2016joint} constructed a Takagi-Sugeno-Kang (TSK) FIS based on joint distribution adaptation (JDA) to simultaneously reduce the difference between the marginal distribution and the conditional distribution of the EEG training sets and test sets. This algorithm can be extended to multi-categorical EEG seizure detection tasks. Furthermore, Jiang \etal \cite{jiang2020eeg} migrated the TSK-FIS to the driver fatigue detection task, and proposed an online multi-view \& transfer TSK-FIS for driver drowsiness estimation. This FIS is self-interpreted, which allows direct tracing of the EEG channels associated with fatigue.
FIS provides human-readable rules that give insights for model refinement and brain signal interpretation. Feng \etal \cite{feng2016joint} developed a Takagi-Sugeno-Kang (TSK) FIS based on joint distribution adaptation (JDA) to simultaneously reduce the difference between the marginal distribution and the conditional distribution of the EEG training sets and test sets. This approach can be extended to multi-categorical EEG seizure detection tasks. Furthermore, Jiang \etal \cite{jiang2020eeg} applied the TSK-FIS to driver fatigue detection, and proposed an online multi-view \& transfer TSK-FIS for driver drowsiness estimation. This FIS is inherently interpretable, enabling direct tracing of the EEG channels associated with fatigue.

\subsubsection{Bayesian System (BS)}
% The BS represents the associations between EEG features and predictions by computing the posterior probability of EEG signal classification. A typical Bayesian model is mathematically related through Bayes' theorem:
The BS uses Bayesian theorem to model the relationship between EEG features and predictions:
\begin{equation}
P(\theta \mid D)=\frac{P(D \mid \theta) P(\theta)}{P(D)},
\end{equation}
% where $\theta$ denotes the model parameters, $D$ represents the observed data (\eg, EEG signals and associated cognitive states), $P(\theta)$ is the prior distribution capturing our prior beliefs about the parameters, $P(D|\theta)$ is the likelihood function indicating the probability of observing the data given the parameters, and $P(\theta|D)$ is the posterior distribution expressing the updated beliefs about the parameters. According to the AI in EEG systems, a Bayesian model can be defined as follows.
where $\theta$ denotes the model parameters, and $D$ represents the observed data (\eg, EEG signals and associated cognitive states). $P(\theta)$ is the prior distribution capturing our prior beliefs about the parameters that relate EEG signals to cognitive states. % This could be based on domain knowledge or expert experience.
$P(D|\theta)$ is the likelihood function that quantifies the probability of observing the EEG data given the model parameters. Depending on the problem, this might involve a linear regression model, a neural network, or any other appropriate models that capture the. $P(\theta|D)$ is the posterior distribution representing the updated beliefs about the model parameters after observing the EEG data.

The interpretability of Bayesian Systems in EEG applications stems from the explicit representation of uncertainty through probability distributions. By analyzing the posterior distribution $P(\theta|D)$, it becomes possible to explain the relationships between EEG signals and cognitive states, as well as the uncertainty associated with the model's predictions.

Qian \etal \cite{qian2016drowsiness} introduced a Bayesian-copula discriminant classifier (BCDC) to study the relationship between drowsiness and nap. The BCDC shows a better understanding of the periodical rhymes of physiological states, and enhances the interpretability of driver alertness. Wu \etal \cite{wu2021separation} proposed a separation and recovery Bayesian method, finding that the predictive emotion features originate from the lateral temporal area and distribute in $\gamma$ and $\beta$ bands.

\if 0
\begin{table*}[h]
\centering
\caption{Summarization of Robustness AI in EEG Systems}
\label{tab2-}
\begin{tabular}{@{}c|c|c@{}}
\hline
Undesirable Factor & Subcategory & Methods and Representative Works \\ \hline
Noise and Artifacts &
  \begin{tabular}[c]{@{}c@{}}External Noise\\ Internal Artifacts\end{tabular} &
  \begin{tabular}[c]{@{}c@{}}Traditional Signal Processing \cite{albera2012ica,kaur2021eeg}\\ Models' Self-Robustness \cite{hussein2019optimized,zhang2021eeg}\end{tabular} \\ \hline
\multirow{2}{*}{Human Variability} & Cross-subject Issues  & TL \cite{li2019multisource}, Robust Feature Extraction \cite{li2022dynamic} \\ \chline
 & Cross-session Issues & TL \cite{lin2019constructing}, Extract Robust Features \cite{yin2017cross,li2022dynamic} \\ \hline
\multicolumn{1}{l|}{\multirow{2}{*}{Data Acquisition Instability}} & Resistance Change & Attention Mechanism \cite{bahador2021reconstruction} \\ \chline
\multicolumn{1}{l|}{}                                                      & Channel Missing \& Broken & Reconstruct Missing Data  \cite{banville2022robust} \\ \hline
Adversarial Attacks                                                        & Evasion, Manipulation     & Adversarial Training \cite{ni2022improving} \\ \hline
\end{tabular}
\end{table*}
\fi

\begin{table*}[t]
  \renewcommand{\arraystretch}{1.1}
  \newcommand{\tabincell}[2]{\begin{tabular}{@{}#1@{}}#2\end{tabular}}
  \caption{Summary of Robust AI in EEG Systems.}
  \label{tab2}
  \centering
  \resizebox{0.95\textwidth}{!}{
  \begin{tabular}{c|c|c}
  \noalign{\hrule height 1.1pt}
  \rowcolor{gray!20}
  \textbf{Undesirable Factors} & \textbf{Subcategory} & \textbf{Methods and Representative Works} \\
  \hline
  \multirow{2}{*}{Noise and Artifacts} & External Noise & Traditional Signal Processing \cite{kaur2021eeg,WAN2022186} \\
  & Internal Artifacts & Models' Self-Robustness \cite{hussein2019optimized,zhang2021eeg} \\
  \hline
  \multirow{2}{*}{Human Variability} & Cross-subject Issues & Transfer Learning  \cite{li2019multisource,wang2023eeg}, Dynamic Domain Adaptation \cite{li2022dynamic} \\
  & Cross-session Issues & Transfer Learning \cite{lin2019constructing,8793187}, Robust Feature Extraction \cite{yin2017cross,li2022dynamic} \\
  \hline
  \multirow{2}{*}{Data Acquisition Instability} & Resistance Change & Attention Mechanism \cite{bahador2021reconstruction,9204431} \\
  & Channel Missing \& Broken & Missing Data Reconstruction \cite{banville2022robust,8982996,8852101,8217773,WAN2022186} \\
  \hline
  Adversarial Attacks & Evasion \& Manipulation & Adversarial Training \cite{ni2022improving,zhang2019vulnerability,liu2021universal,feng2021saga} \\
  \noalign{\hrule height 1.1pt}
  \end{tabular}
  }
\end{table*}

\subsection{Discussion on Interpretable AI in EEG Systems}
Selecting an appropriate interpretability method is crucial for understanding the decision-making mechanisms of AI models in EEG systems. Each has its own strengths and limitations, which we will explore in more detail. The comparative overview of these methods is shown in Table \ref{tab:cmp_interpre}.

% Backpropagation-based methods are a popular subset of interpretability techniques, and they include approaches such as LRP, DeepLIFT, CAM, and Grad-CAM. These are all post-hoc techniques, which means that they are applied after the model has made its predictions to identify the time-frequency patterns and electrode regions that were influential in the decision-making process. Backpropagation-based methods can offer valuable insights into the model's behavior by illuminating the inner workings of its hidden layers. However, these techniques are fundamentally limited by their dependence on gradient information, which is only sometimes reliable or available. Furthermore, they are model-specific, meaning that the interpretations they generate are tied to the particular model they are applied to, constraining their broader applicability.
Backpropagation-based methods represent a popular subset of interpretability techniques, encompassing methods like LRP, DeepLIFT, CAM, and Grad-CAM. These are post-hoc methods, which mean that they are applied after the model has made its predictions to identify the time-frequency patterns and electrode regions that were influential in the decision-making process. They can offer valuable insights into the model's behavior by illuminating the inner workings of its hidden layers. However, they rely on gradient information which may be unavailable. Moreover, their model-specific nature limits broader applicability.

% In contrast, perturbation-based methods such as LIME and SHAP provide greater flexibility because they are model independent. These methods deliver comprehensive insights into a model's predictive behavior by creating local surrogate models that approximate the original model's behavior in a particular instance's neighborhood. This can reveal how different features contribute to a prediction, which can be particularly useful in EEG systems analysis, where it is often crucial to understand which brain regions and time frequency features are most important. However, these methods have their drawbacks. They can be computationally intensive, especially for complex models or large datasets. Moreover, since these methods involve fitting local models, they may be prone to overfitting, which can lead to misinterpretations.
In contrast, perturbation-based methods such as LIME and SHAP provide greater flexibility as they are model-independent. These methods explain the model's predictive behavior by creating local surrogate models that approximate the original model's behavior in a particular instance's neighborhood. This can reveal how different features contribute to a prediction, which is valuable in EEG systems analysis to understand crucial brain regions and time-frequency features. However, these methods can be computationally intensive, especially for complex models or large datasets. Moreover, fitting local models may raise overfitting concerns, potentially leading to misinterpretations.

% Lastly, we focus on rule-based methods such as RF, FIS and BS. These methods are designed to be interpretable from the outset, using specific logic rules or mathematical statements to enhance transparency. This allows them to provide direct insights into how different features contribute to the model's predictions, making it easier for users to trust and understand the model. However, these methods also come with their own set of challenges. They can oversimplify complex EEG systems by reducing them to a set of rules or equations, which may not fully capture the intricate interactions and nonlinearities in the data. Moreover, they often require domain expertise to interpret correctly, and there is a need for careful evaluation of trade-offs between interpretability and performance, as these methods may not consistently deliver the highest predictive accuracy.
Rule-based methods like RF, FIS, and BS prioritize interpretability by using logic rules or mathematical statements. Their transparency allows direct insight into feature contributions, promoting trust and understanding. However, they may oversimplify complex EEG systems by reducing them to incomplete rules.  Moreover, accurate interpretation often requires domain expertise, and a balance between interpretability and performance evaluation is necessary, as these methods might not consistently yield the highest predictive accuracy.

% In conclusion, selecting an appropriate interpretability method involves careful consideration of the trade-offs among the specificity of backpropagation-based methods, the computational cost and risk of overfitting associated with perturbation-based methods, and the potential for oversimplification by rule-based methods. The chosen method should match the specific needs and constraints of the EEG system under investigation, the computational resources available, the required level of interpretability, and the expertise of the users.
In conclusion, selecting an appropriate interpretability method involves the following aspects: the specificity of backpropagation-based method, the computational cost and overfitting risks of perturbation-based methods, and the incompleteness of rule-based methods. The selection should match the specific needs and constraints of the EEG system, the computing resources available, the required level of interpretability, and the expertise of the users.

\section{Robust AI in EEG systems}
% The robustness of EEG systems refers to their ability to consistently and accurately perform their designated tasks, even when faced with unexpected or complex conditions. This includes being able to handle variations in the data input, such as changes in the brain activity patterns or noise in the signal, as well as being resistant to external factors such as power fluctuations or hardware malfunctions. Ensuring robustness in EEG systems is essential for their reliability and effectiveness in real-world applications, such as diagnosing and treating brain disorders or developing brain-computer interfaces. 
% AI robustness refers to the ability of AI models to consistently and accurately perform their designated tasks when faced with unexpected conditions.  The robustness of EEG systems is essential for their practical application. Foremost, robust EEG systems resist uncontrollable disturbances in EEG signals from the sampling process to the signals' components. Robustness methods allow EEG systems to adapt to, for example, changes in subjects' brain activity patterns, mitigate interference from noise in the environment, and fill in channel deficits caused by fluctuations in electrode resistance. In addition, the robustness approach can significantly improve the model's overall accuracy in practical applications. For example, EEG systems are frequently used to diagnose and treat brain diseases, and the models' accuracy directly determines whether the EEG system's predictions are informative for the physician's diagnosis.
AI robustness refers to the ability of AI models to consistently and accurately perform their designated tasks when faced with unexpected conditions. For robust AI in EEG systems, it means effectively countering uncontrollable disturbances across the entire spectrum of EEG signal processing, spanning from the signal sampling phase to the signal input phase. Specifically, the robust AI should adapt to changing brain activity patterns, resist environmental noise, and fill channel gaps from electrode resistance fluctuations. Moreover, model robustness ensures model accuracy in practical applications. For example, physicians rely on precise EEG diagnostics to inform brain disease treatment. Without robust AI providing accurate predictions, the EEG systems lack diagnostic value.

In this section, we classify the undesirable factors that affect EEG systems into four categories: noise and artifacts, human variability, data acquisition instability and adversarial attacks. Based on these four categories, we elaborate the techniques that can alleviate the adverse effects of undesirable factors and improve the robustness of EEG systems. We summarize the robust AI and the representative works in Table~\ref{tab2}.

%\subsection{Traditional Robustness }
\subsection{Noise and Artifacts in Signals}
Brain activity measurement, especially through the EEG signals, is often susceptible to external noises from various sources (\eg, electromagnetic interference) and internal artifacts from the human body (\eg, muscle movements and eye blinks). These factors distort or interfere with EEG signals, impacting the accuracy and reliability of the EEG system. To address this issue, two types of methods have been developed to minimize the effects of noise and artifacts in EEG signals.

\subsubsection{Signal Processing}
The first type involves traditional signal processing denoising techniques, relying on filtering algorithms developed from prior knowledge to separate and remove noise efficiently.
For instance, Kaur \etal \cite{kaur2021eeg} compared two signal denoising techniques based on discrete wavelet transform (DWT) and wavelet packet transform (WPT) combined with variational mode decomposition (VMD).
The VMD first decomposes the signals into diverse components, and then the DWT and WPT are used to denoise the artifactual components.
The WPT with VMD provides a more refined frequency decomposition, facilitating better noise separation and artifact removal compared to its DWT counterpart.

\subsubsection{Learning-based Denoising}
The second type involves the use of modular adaptive denoising techniques in EEG systems, which rely on specialized network structures to automatically denoise the signals. Hussein \etal \cite{hussein2019optimized} utilized a long short-term memory (LSTM) network to leverage the temporal dependencies in the time series EEG data, and the LSTM can be expressed as

\begin{equation}
\begin{aligned}
f_t &= \sigma(W_f [h_{t-1}, x_t] + b_f), \\
i_t &= \sigma(W_i [h_{t-1}, x_t] + b_i), \\
o_t &= \sigma(W_o [h_{t-1}, x_t] + b_o), \\
\tilde{c}_t &= \tanh(W_c [h_{t-1}, x_t] + b_c), \\
c_t &= f_t * c_{t-1} + i_t * \tilde{c}_t, \\
h_t &= o_t * \tanh(c_t)
\end{aligned}
\end{equation}
% \begin{equation}
% \resizebox{0.55\columnwidth}{!}{$
% \begin{split}
% f_t = \sigma(W_f\cdot[h_{t-1}, x_t] + b_f), \\
% i_t = \sigma(W_i\cdot[h_{t-1}, x_t] + b_i), \\
% \tilde{C}_t = \tanh(W_C\cdot[h_{t-1}, x_t] + b_C), \\
% C_t = f_t * C_{t-1} + i_t * \tilde{C}_t, \\
% o_t = \sigma(W_o\cdot[h_{t-1}, x_t] + b_o), \\
% h_t = o_t * \tanh(C_t)
% \end{split}
% $}
% \end{equation}
where $f_t$, $i_t$ and $o_t$ represent the forget, input and output gates at timestamp $t$, $W_{\cdot}$ and $b_{\cdot}$ refer to the weight and bias for the respective gates, $c_t$ and $\tilde{c}_t$ denote the cell state and the cell input activation, $\sigma(\cdot)$ denotes the Sigmoid function, $*$ denotes the element-wise product, and
$h_t$ is the output hidden state. 
The robustness of \cite{hussein2019optimized} lies in the LSTM's ability to learn long-term dependencies and capture relevant temporal features, thus improving signal-to-noise ratio (SNR) and reducing artifacts.

Building on the inception-time network backbone, Zhang \etal \cite{zhang2021eeg} proposed an end-to-end framework that takes raw EEG signals as input, eliminating the need for complex signal preprocessing. This noise-insensitive method can capture robust features of motor imagery (MI) tasks and effectively eliminate noise interference. The inception-time network is described as

\begin{equation}
F(x) = \sum_{i=1}^{N} w_i f_i(x),
\end{equation}
where $F(x)$ represents the output of the inception layer, $w_i$ denotes the weights, $f_i(x)$ and $N$ refers to the $i$-th convolutional layer and the total layer numbers. The robustness of this method mainly stems from the network's ability to learn hierarchical representations of EEG signals and adaptively select relevant features, enhancing the noise suppression capabilities and improving the overall signal quality.

% Both traditional signal processing denoising techniques and learning-based denoising methods contribute significantly to enhancing the robustness of EEG data. The mathematical models of these approaches, such as the wavelet transforms and neural network architectures, demonstrate the sophisticated nature of these techniques in mitigating noise and artifacts in EEG signals.

% By focusing on improving signal quality and robustness, researchers can obtain more accurate and reliable data for various applications, including the diagnosis and treatment of neurological disorders, brain-computer interfaces, and cognitive research. As technology continues to advance, it is expected that further improvements in noise reduction and artifact removal methods will be made, further enhancing the quality of EEG data and its potential impact on various fields of study.

Both traditional signal processing denoising techniques and learning-based denoising methods contribute significantly to enhancing the robustness of EEG data. 
By focusing on improving signal quality and robustness, it is expected that more accurate and reliable data can be obtained for various applications, including the diagnosis and treatment of neurological disorders, brain-computer interfaces and cognitive research.

\subsection{Human Variability}
% EEG signal variations across different subjects and states for the same subject present significant challenges for cross-subject and cross-session EEG systems. These variations stem from differences in brain anatomy, brain function, and other individual characteristics, while changes in mental state, fatigue level, or recording environment for the same subject can also result in significant disparities in EEG signals. To mitigate the influence of these variations, models must extract stable features from cross-session/subject EEG signals performing the same task.

% One common approach to address the EEG variations caused by human variability is transfer learning (TL). For cross-subject EEG, Li \etal \cite{li2019multisource} propose a multisource TL method, which applies transfer mapping to reduce the difference between known and new subjects, by minimizing the following divergence:
Variations in EEG signals across different subjects and states pose significant challenges for cross-subject and cross-session EEG systems. These variations mainly arise from the differences in brain anatomy, brain function, and other individual characteristics. Additionally, changes in mental state, fatigue, or recording conditions for the same subject can also lead to substantial disparities in EEG data. To mitigate such impact, AI models need to extract stable features from EEG signals across sessions and subjects.

One common approach to address the EEG variations caused by human variability is transfer learning (TL). For cross-subject EEG, Li \etal \cite{li2019multisource} proposed a multi-source TL method, which utilizes transfer mapping to reduce the difference between known and new subjects, by minimizing the following divergence:

\begin{equation}
D(\mathcal{S}, \mathcal{T}) = \sum_{i=1}^{n} ||\mathbf{w}_{i}^{\mathcal{S}} - \mathbf{w}_{i}^{\mathcal{T}}||_{2}^{2},
\end{equation}
where $\mathcal{S}$ and $\mathcal{T}$ denote the source domain and target domain, $\mathbf{w}_{i}^{\mathcal{S}}$ and $\mathbf{w}_{i}^{\mathcal{T}}$ represent the weights in the corresponding domains, and $||\cdot||_2$ denotes the $\ell_2$-norm. 

For cross-session EEG, Lin \etal \cite{lin2019constructing} proposed a robust principal component analysis (RPCA)-embedded TL approach, aiming to generate a personalized cross-session model with less labeled data while alleviating intra-session and inter-session differences. The loss function of the TL is given by

\begin{equation}
\min_{\mathbf{L},\mathbf{E}} ||\mathbf{L}||_* + \lambda ||\mathbf{E}||_1 \quad \text{s.t.} \quad \mathbf{X} = \mathbf{L} + \mathbf{E},
\end{equation}
where $\mathbf{X}$ is the data matrix of EEG signals, $\mathbf{L}$ is the low-rank matrix, $\mathbf{E}$ is the sparse error matrix, and $\lambda$ is the regularization parameter.  $\|\cdot\|_*$ denotes the matrix nuclear norm, which is the sum of the singular values, and $\|\cdot\|_1$ denotes the $\ell_1$-norm, which is the sum of the absolute values of entries.

% Additionally, self-adaptive methods are widely used to extract robust EEG features. Li \etal \cite{li2022dynamic} emphasize the importance of aligning the EEG data within the same emotion categories for generalizable and discriminative features. They propose the dynamic domain adaptation (DDA) algorithm, where global and local divergences are handled by minimizing global and local subdomain discrepancies:
Apart from the TL methods, self-adaptive methods are also used to extract robust EEG features to deal with the human variability. Li \etal \cite{li2022dynamic} emphasized the importance of aligning EEG data within the same emotion class for generalizable and discriminative features. They proposed a dynamic domain adaptation (DDA) algorithm, where global and local divergences are handled by minimizing their subdomain discrepancies:

\begin{equation}
\min_{\mathbf{F}_{s}, \mathbf{F}_{t}} \!\mathcal{L}(\mathbf{F}_{s}, \mathbf{F}_{t}) \!=\! \sum_{c=1}^{C} D_{c}(\mathbf{F}_{s}, \mathbf{F}_{t}) \!+\! \alpha \sum_{c=1}^{C} \!D_{\text{local}}(\mathbf{F}_{s}^{c}, \mathbf{F}_{t}^{c}),
\end{equation}
% where $\mathbf{F}_{s}$ and $\mathbf{F}_{t}$ denote the feature representations of the source domain and target domain.  $C$ is the number of classes, and $\alpha$ is the regularization parameter.
where $\mathbf{F}_s$ and $\mathbf{F}_t$ represent the feature representations in the source domain and target domain, $\mathbf{F}_s^c$ and $\mathbf{F}_t^c$ represent the feature representations for the $c$-th class in two domains, $C$ indicates the total number of classes, $\alpha$ denotes the regularization parameter, $D_c$ computes the global divergence between the two domains for class $c$, and $D_{\text {local }}$ computes the local divergence for each class between two domains. The DDA intends to harmonize the feature representations between the source and target domains.

Furthermore, motivated by the effectiveness of deep learning approaches for stable feature abstractions at higher levels, Yin and Zhang \cite{yin2017cross} developed an adaptive stacked denoising autoencoder (SDAE) to extract cross-session EEG features.
% Within this framework, the weights of the first hidden layer, directly connected to the input layer, undergo iterative updates. This process accounts for the shifts in the statistical properties of EEG power features observed over consecutive days. Consequently, the SDAE model is endowed with the proficiency to capture a precise EEG data distribution at a more elevated level.
Within this framework, the weights of the first hidden layer are directly connected to the input layer and are updated iteratively. This process accounts for the shifts in the statistical properties of EEG power features observed over consecutive days. Consequently, the SDAE model is endowed with the proficiency to capture a precise EEG data distribution at a high level.

\subsection{Data Acquisition Instability}
% Data acquisition instability refers to the unstable connection between the EEG acquisition equipment and the subject, and it leads to the loss of EEG channels. One cause of unstable connection is the hardening of the glue connecting the electrodes to the scalp over time, as this increases the resistance of the electrodes. Also, sweating on the subject's scalp would have a similar effect. In the above two cases, the electrode impedance changes are difficult to detect, so it causes undetectable channel loss in EEG signals. Foremost, the solution to this problem is to find out the missing channels.

% Banville \etal \cite{banville2022robust} propose dynamic spatial filtering (DSF), a multi-head attention mechanism that can be used to focus on good channels and ignore bad ones. The DSF computes the attention weights for each channel as follows:
Data acquisition instability refers to the unstable connection between the EEG acquisition equipment and the subject, resulting in the loss of EEG channels. One factor that leads to this instability is the hardening of the glue connecting the electrodes to the scalp over time, thus increasing the resistance of the electrodes. Besides, sweating on the subject's scalp can have a similar effect. In the above two cases, the changes of electrode impedance are difficult to detect, so they cause undetectable channel loss in EEG signals. The primary solution to this issue is to identify the missing channels.

Banville \etal \cite{banville2022robust} proposed dynamic spatial filtering (DSF), a multi-head attention mechanism that focuses on good channels and ignore bad ones. The DSF computes the attention weights for each channel as follows:

\begin{equation}
\alpha_i = \frac{\exp(\mathbf{W}_a \mathbf{X}_i)}{\sum{j=1}^{N} \exp(\mathbf{W}_a \mathbf{X}_j)},
\end{equation}
where $\alpha_i$ represents the attention weight for the $i$-th channel, $\mathbf{W}_a$ is the learnable attention matrix, $\mathbf{X}_i$ is the feature vector for the $i$-th channel, and $N$ is the total number of channels.

Estimating and reconstructing the missing channels is another promising way to address this issue. Bahador \etal \cite{bahador2021reconstruction} estimated and reconstructed the data segments in missing channels based on the information near the missing segments. They used a linear weighted interpolation method

% \begin{equation}
% \mathbf{Y}{\text{miss}} = \sum{i=1}^{N} w_i \mathbf{Y}_{\text{neigh}, i},
% \end{equation}
\begin{equation}
\mathbf{Y}_m = \sum{i=1}^{N} w_i \mathbf{Y}_{n, i},
\end{equation}
where $\mathbf{Y}_m$ is the estimated missing channel data, $\mathbf{Y}_{n, i}$ represents the $i$-th neighboring channel data, $w_i$ is the corresponding weight, and $N$ is the total number of neighboring channels.

\subsection{New Emerging: Adversarial Attacks}
% Adversarial attacks on EEG systems have become a growing concern in neuroscience and cybersecurity. EEG systems are widely used in medical-related fields, such as pathological diagnosis, control of bionic prosthetic limbs and communication of severely disabled individuals (\eg, amyotrophic lateral sclerosis patients). Since these scenarios involve patient privacy and safety, if the EEG systems cannot resist adversarial attacks, they may cause severe medical accidents.
Adversarial attacks on EEG systems have become a growing concern in neuroscience and cybersecurity. EEG systems are widely used in medical-related fields, including pathological diagnosis, control of bionic prosthetic limbs, and communication of severely disabled individuals (\eg, amyotrophic lateral sclerosis patients). Since these scenarios involve patient privacy and safety, the vulnerability of EEG systems to adversarial attacks may cause severe medical accidents.

% Adversarial attacks on EEG systems usually consist of evasion and manipulation. Evasion refers to making EEG systems unable to make correct judgments by creating misleading EEG signals \cite{zhang2021tiny, feng2021saga}. For example, an attacker could interfere with users' MI EEG to cause their bionic prosthetic to lose control, which could easily cause serious injury to the user or bystanders. As for the manipulation, it means simulating the user's EEG to manipulate systems. Attackers can mislead EEG systems to signal an incorrect intention on the part of the person with a disability. In such a scenario, EEG systems may leak the personal information of the individual with a disability or initiate unwanted financial transactions.

% However, it is difficult to apply the adversarial attacks to EEG systems in a real-word scenario, and thus the defensive methods against adversarial attacks, such as adversarial training, are only beginning to be investigated \cite{ni2022improving}. Adversarial training is a technique that incorporates adversarial examples into the training process to enhance the robustness of the model. Given an EEG input $\mathbf{x}$ and its corresponding label $y$, the adversarial training aims to minimize the loss function $\mathcal{L}$ under adversarial perturbations:

Adversarial attacks on EEG systems typically consist of evasion and manipulation. Evasion involves crafting misleading EEG signals to cause the EEG systems to yield incorrect predictions \cite{zhang2021tiny, feng2021saga}. For example, an attacker could interfere with users' EEG in a MI task to make bionic prosthetic lose control, which may potentially injure the users or bystanders. As for the manipulation, it means simulating the user's EEG to deceive EEG systems into misinterpreting the user's intentions. In such a scenario, EEG systems may leak individual's personal information or initiate unauthorized financial transactions.

However, real-world adversarial attacks on EEG systems are rather difficult, and thus the defensive methods against these attacks have just begun to be investigated \cite{ni2022improving}. Adversarial training is a defensive technique that incorporates adversarial examples into the training process to enhance model robustness. Given an EEG input $\mathbf{x}$ and its corresponding label $y$, the adversarial training aims to minimize the loss function $\mathcal{L}$:

\begin{equation}
\min_{\theta} \mathbb{E}{(\mathbf{x}, y)}[\mathcal{L}(f(\mathbf{x}_{\mathrm{adv}}; \theta), y)],
\end{equation}
where $\mathbf{x}_{\text{adv}}$ denotes the adversarial example, $f$ represents the model with model parameter $\theta$, and the expectation is taken over the distribution of training data $(\mathbf{x}, y)$. By minimizing the loss function under adversarial perturbations, the model's resilience to adversarial attacks is improved, ensuring the safety and privacy of patients using EEG systems.

\subsection{Discussion on Robust AI in EEG Systems}
% Understanding and mitigating the factors impacting the robustness of AI-based Electroencephalography (EEG) systems is critical in the rapidly developing field of neural engineering and AI. These factors, ranging from noise and artifacts to human variability, data acquisition instability, and adversarial attacks, can critically degrade the performance and reliability of these AI-based systems. Each of these factors originates from unique sources and presents distinct challenges, necessitating a comprehensive, multi-pronged approach to address and mitigate them effectively.

% Noise and artifacts from external and internal physiological sources significantly challenge EEG signal acquisition and interpretation quality. Externally, noise sources can be as diverse as electromagnetic interference from nearby electronic devices to more trivial sources such as muscle movements, all contributing to the degradation of the EEG signal quality. Internally, sources of noise and artifacts can include heart rhythms, eye movements, and other biological phenomena. This noise can seriously compromise the signal-to-noise ratio of the EEG recordings, making them more challenging to analyze and interpret. Mitigation strategies for such noise typically involve stringent experimental protocols and robust signal processing algorithms that filter out noise without compromising the integrity of the underlying neural signal.
Understanding and mitigating the factors that impact the robustness of AI-based EEG systems is vital in the burgeoning field of neural engineering and AI. These factors, including noise and artifacts, human variability, data acquisition instability and adversarial attacks, can critically degrade the performance and reliability of the EEG systems. Each factor arises from unique sources and presents distinct challenges, necessitating a comprehensive and multi-pronged approach to address them effectively.

Noise and artifacts from external and internal physiological sources significantly impede the quality of EEG signal acquisition and interpretation. Externally, diverse noise sources like electromagnetic interference or muscle movements, detrimentally affect the EEG signal fidelity. Internally, noise and artifacts from heart rhythms, eye movements and other biological phenomena can compromise the SNR of the EEG recordings, making them more difficult to analyze. Mitigating such issues requires stringent experimental protocols and robust signal processing algorithms that filter out noise without compromising the integrity of the underlying neural signals.

% Similarly, the inherent variability among human subjects presents another challenge to AI-based EEG systems. This variability can manifest in numerous ways, including differences in skull thickness and scalp conductivity, cognitive states, and other biological factors. Furthermore, temporal variations, such as changes in a person's mental state or fatigue level, can also impact the EEG signals. Therefore, designing AI-based EEG systems that can generalize across such inter- and intra-individual differences is paramount. Solutions include developing sophisticated machine learning models that account for individual variations or implementing adaptive algorithms capable of adjusting to temporal variations.
The inherent variability among human subjects poses another challenge to AI-based EEG systems. This variability can manifest in numerous ways, through the differences in skull thickness and scalp conductivity, cognitive states, and other biological factors. Furthermore, temporal variations, such as changes in a person's mental state or fatigue level, can also impact the EEG signals. Therefore, designing AI-based EEG systems that can generalize across inter- and intra-individual differences is paramount. Solutions involve developing sophisticated machine learning models that account for individual variations or implementing adaptive algorithms capable of adjusting to temporal variations.

% Data acquisition instability is another significant factor impacting the robustness of AI-based EEG systems. This instability can result from technical issues such as changes in electrode-skin resistance, missing data due to broken or disconnected channels, or malfunctioning recording devices. These issues lead to data loss or degradation, significantly affecting the quality and interpretability of the data. As such, mitigation strategies often target hardware and software improvements, including more robust EEG devices, improved electrode design and materials, and better error detection and correction algorithms.
Data acquisition instability is another significant factor impacting the robustness of AI-based EEG systems. This instability can stem from technical issues such as changes in electrode-skin resistance, missing data due to broken or disconnected channels, or malfunctioning recording devices. These issues lead to data loss or degradation, significantly hampering the quality and interpretability of the EEG data. Therefore, solutions typically focus on the improvements in hardware and software, including more stable EEG devices, improved electrode design and materials, and more efficient error detection and error correction algorithms.

% Lastly, adversarial attacks substantially threaten the security and integrity of AI-based EEG systems. These attacks typically involve intentionally manipulating input data to exploit vulnerabilities in the AI model, causing it to make incorrect predictions or classifications. Adversarial attacks can be especially damaging in the context of AI-based EEG systems, given the sensitive nature of the data and the potential for misuse. Proactive defenses are necessary to protect against such threats, including improving model robustness through adversarial training, implementing rigorous data integrity checks, and developing robust cybersecurity measures.
Lastly, adversarial attacks substantially threaten the security and integrity of AI-based EEG systems. These attacks often exploit the vulnerabilities in AI models by intentionally manipulating input data, leading to incorrect predictions or classifications. Proactive defenses are necessary to resist such threats, including improving model robustness through adversarial training, implementing rigorous data integrity checks, and developing robust cybersecurity measures.

% Understanding the relationships and differences between these factors is crucial for developing comprehensive strategies to enhance the performance, reliability and security of AI-based EEG systems. By identifying and addressing these distinct challenges, researchers and engineers can enhance the robustness of these systems, facilitating their adoption in a broader range of real-world applications. At the same time, they can ensure that these systems are secure and resilient, protecting the data and privacy of the individuals who use them. This comprehensive approach is vital to unlocking the full potential of AI-based EEG systems and propelling the field forward.

\section{Future Directions}
% Interpretability and robustness techniques offer a promising future for building better EEG systems. With an eye on the future, we propose directions for updating interpretability and robustness techniques in EEG systems.
Interpretability and robustness techniques offer a promising future for building better EEG systems. Despite much success, there are still some unresolved problems worthy of in-depth study. Therefore, we discuss a few promising directions.

\subsection{Future Directions for Interpretable AI in EEG Systems}

\subsubsection{Prior Human Knowledge}

A practical limitation of existing interpretable EEG systems is the inability to integrate prior information. For EEG systems, prior information refers to the established physiological principles. Attribution methods only provide correlations between features and predictions, but models may still pay attention to the features we do not want them to learn. We expect the interpretable EEG systems in real-world scenarios to jointly learn the relationship between prior information and feature importance, ensuring that the explanations rely on features predicted to be essential.

% The DAPr framework can be formalized as follows:

% \begin{equation}
% \min_{\theta} \mathbb{E}_{(\mathbf{x}, y)}[\mathcal{L}(f(\mathbf{x}; \theta), y)] + \lambda \mathcal{R}(\theta),
% \end{equation}

% where $\mathcal{L}$ is the loss function, $f$ is the prediction model, $\theta$ represents the model parameters, and $\mathcal{R}(\theta)$ is the regularization term that incorporates the prior knowledge. The parameter $\lambda$ controls the trade-off between the prediction accuracy and the alignment with prior knowledge.

% Similarly, for instance, in neuroscience, the seizure has been proven to be caused by sudden abnormal discharges of neurons in the temporal lobe region of the brain. While in its diagnosis, EEG systems are easily susceptible to interference from noise generated by the patient's muscle movements. In this case, feeding the seizure region as prior knowledge into models can help the AI focus on features to receive constraints, allowing systems' predictions to align with doctor knowledge. By incorporating the prior information into the regularization term, the model can be guided to concentrate on the relevant brain regions and ignore the irrelevant noises, thus improving the robustness and interpretability of the EEG system.
Weinberger \etal \cite{weinberger2020learning} proposed a deep attribution prior (DAPr) framework that acquires prior knowledge and constraints the model via a prior model.
Similarly, for instance, in neuroscience, seizures are known to be caused by sudden abnormal discharges of neurons in the temporal lobe. While the EEG systems in seizure diagnosis are easily interfered with the noise generated by the patient's muscle movements. In this case, taking the seizure region as prior knowledge could help models focus on relevant features and constrain predictions to align with medical knowledge.

% \subsubsection{High-dimensional Feature Interpretation}
%%%原版%%%%%%%%%%%%%%%%%%%%%%%%%%%%%%%%%%%%%
% The existing interpretable methods in EEG systems mainly reveal the features' contributions to the prediction, but they lack insight into why features are assigned specific contributing values. Providing dynamic feature descriptions rather than only linear relationships between features and predictions can be a promising way to reveal the inner logic of model predictions. For instance, Zhang \etal \cite{zhang2018interpretable} design special loss for each conv-layer and set them to focus on certain areas of the input image. Sabour \etal \cite{sabour2017dynamic} propose capsule networks that parse the entire object into a parsing tree of capsules by a dynamic routing mechanism, and each capsule may encode a specific meaning of input data. 

% We expect interpretable AI in future EEG systems to gain insight from the above two works to provide hidden semantics to explain feature correlations. For instance, in the MI task, high-dimensional features could be interpreted as concurrency between low-latitude motor cortex signal features and visual cortex signal features. Alternatively, with models' hidden semantics, we can know how models' attention is drawn to noises if high-dimensional features can be interpreted as similarities between noise and essential features.
%%%%%%%%%%%%%%%%%%%%%%%%%%%%%%%%%%%%%%%%%%%%

\subsubsection{High-dimensional Feature Interpretation}
% The existing interpretable methods in EEG systems mainly reveal the features' contributions to the prediction, but they lack insight into why features are assigned specific contributing values. Providing dynamic feature descriptions rather than only linear relationships between features and predictions can be a promising way to reveal the inner logic of model predictions. For instance, Zhang \etal \cite{zhang2018interpretable} design special loss for each conv-layer and set them to focus on certain areas of the input image. Sabour \etal \cite{sabour2017dynamic} propose capsule networks that parse the entire object into a parsing tree of capsules by a dynamic routing mechanism, and each capsule may encode a specific meaning of input data.
The existing interpretable methods in EEG systems mainly reveal the contributions of features to predictions, but lack insight into why features are assigned specific contributing values. Providing dynamic feature descriptions, rather than only linear relationships between features and predictions, can be a promising way to reveal the inner logic of model predictions. For example, Zhang \etal \cite{zhang2018interpretable} designed special loss for each convolutional layer, instructing them to focus on certain regions within the input image. Sabour \etal \cite{sabour2017dynamic} proposed capsule networks to parse the entire object into a parsing tree of capsules by a dynamic routing mechanism, in which each capsule may encode a specific meaning of input data.

% In capsule networks, the dynamic routing algorithm can be represented as follows:

% \begin{equation}
% c_{ij} = \frac{\exp(b_{ij})}{\sum_{k} \exp(b_{ik})},
% \end{equation}

% where $c_{ij}$ is the coupling coefficient that determines the contribution of capsule $i$ to capsule $j$, and $b_{ij}$ is the log prior probability that capsule $i$ should be coupled to capsule $j$.

% We expect interpretable AI in future EEG systems to gain insight from the above two works to provide hidden semantics to explain feature correlations. For instance, in the MI task, high-dimensional features could be interpreted as concurrency between low-latitude motor cortex signal features and visual cortex signal features. Alternatively, with models' hidden semantics, we can know how models' attention is drawn to noises if high-dimensional features can be interpreted as similarities between noise and essential features. By incorporating dynamic routing algorithms or specialized loss functions into EEG systems, we can guide the model to focus on specific semantic features and enhance the robustness of the system while maintaining interpretability in line with academic standards.
We expect interpretable AI in future EEG systems to gain insight from the above two works to provide hidden semantics to explain feature correlations. For instance, in the MI task, high-dimensional features could be interpreted as concurrency between low-latitude motor cortex signal features and visual cortex signal features. Alternatively, with models' hidden semantics, we can know how models' attention is drawn to noises if high-dimensional features can be interpreted as similarities between noise and essential features. By incorporating dynamic routing algorithms or specialized loss functions into EEG systems, we can guide the model to focus on specific semantic features, and enhance the robustness of the system while maintaining interpretability.

\subsection{Future Directions for Robust AI in EEG Systems}
% In terms of the future robustness of EEG systems, we give several highly promising points:
\subsubsection{Artificial Synthetic Data for Large Models}
%%%%%%%%%原版%%%%%%%%%%%%%%%%%%%%%%%%%%%%%%
% Large-models have been used in NLP and CV, which achieves impressive performance with good robustness. This is achieved by training with large amounts of data and thus naturally performing well against anomalies. However, there has been little work on applying large-models to EEG systems because there is a paucity of EEG data.

% Some existing works utilize traditional generative models, such as GANs \cite{abdelfattah2018augmenting,fahimi2020generative}, to synthesize new EEG data artificially to augment the EEG data. However, the performance of these efforts is very mediocre due to the lack of proper EEG generation mechanisms. In the future, better models will need to be developed to synthesize EEG data. Besides, EEG-oriented data augmentation based on signal processing or adversarial examples is also a potential direction.  
% Once sufficient EEG data is available, we can apply the large-model approach to EEG systems to make them more robust.
%%%%%%%%%%%%%%%%%%%%%%%%%%%%%%%%%%%%%%%%%%%%
% Large-models have been used in NLP and CV, which achieves impressive performance with good robustness. This is achieved by training with large amounts of data and thus naturally performing well against anomalies. However, there has been little work on applying large-models to EEG systems because there is a paucity of EEG data. Some existing works utilize traditional generative models, such as GANs \cite{abdelfattah2018augmenting,fahimi2020generative}, to synthesize new EEG data artificially to augment the EEG data.
Large models have been used in NLP and CV, demonstrating impressive performance coupled with robustness. This is achieved by training with large amounts of data and thus naturally performing well against anomalies. However, there has been little work on applying large models to EEG systems due to the scarcity of available EEG data. Some existing works utilize traditional generative models, such as GANs \cite{abdelfattah2018augmenting, fahimi2020generative}, to artificially synthesize new EEG data for data augmentation.

% The objective function of a GAN can be expressed as:

% \begin{equation}
% \resizebox{0.95\columnwidth}{!}{$
% \displaystyle\min_{G} \displaystyle\max_{D} \Big(\mathbb{E}_{\mathbf{x} \sim p(\text{data})}\big[\log D(\mathbf{x})\big] + \mathbb{E}_{\mathbf{z} \sim p(\text{noise})}\big[\log\big(1 - D(G(\mathbf{z}))\big)\big]\Big),
% $}
% \end{equation}

% where $G$ is the generator network, $D$ is the discriminator network, $p(\text{data})$ is the distribution of real EEG data, and $p(\text{noise})$ is the distribution of random noise. 
% However, the performance of these efforts is very mediocre due to the lack of proper EEG generation mechanisms. In the future, better models will need to be developed to synthesize EEG data. Besides, EEG-oriented data augmentation based on signal processing or adversarial examples is also a potential direction. Once sufficient EEG data is available, we can apply the large-model approach to EEG systems to make them more robust. For instance, we could use transfer learning to adapt large pre-trained models to the specificities of EEG data. By incorporating large-models and advanced data augmentation techniques, we can enhance the robustness of EEG systems while maintaining interpretability in line with academic standards.
However, the performance of these works is rather limited because of a lack of proper EEG generation mechanisms. In the future,  it is imperative to develop more sophisticated models for EEG data synthesis. Besides, EEG-oriented data augmentation based on signal processing or adversarial examples is also a potential direction. Once sufficient EEG data is available, we can apply the large models to EEG systems to make them more robust. For instance, we could use transfer learning to adapt large pre-trained models to the specificities of EEG data.

\subsubsection{Decoupling of EEG Signals for Robust Feature}
% EEG signals contain various information, covering identity information (which can be used for personal identification) and task-related information (MI, emotion recognition, \etc). The above information is highly coupled and interferes with each other.
% EEG signals contain a variety of information, including identity information of subjects and task-related information (MI, emotion recognition, \etc).
% On the one hand, the identity information in EEG signals is more difficult to forge than other biometric information (\eg, face, iris, and fingerprint), so it can be used for more reliable identity recognition. 
% On the other hand, identity information is also a kind of noise that affects the performance of other tasks. Therefore, removing the identity information from the EEG signals can make the EEG systems more robust, immune to the changes of subjects, and thus suitable for cross-subject scenarios.
EEG signals contain diverse information, including subject identity and task-related information (\eg, MI and emotion recognition).
On the one hand, the identity information in EEG signals is more difficult to forge than other biometric information (\eg, face, iris and fingerprint), so it can be used for more reliable identity recognition. 
On the other hand, the identity information is also a kind of noise that affects the performance of other tasks. Therefore, disengaging the identity information from the EEG signals can make the EEG systems more resilient to subject variations, thus enhancing the robustness of EEG systems and allowing for better cross-subject applications.

However, various types of information in EEG signals are highly coupled and interfere with each other, which hinders their applications. Thus, how to decouple the EEG Signals for designing robust features is a promising direction.

% Inspired by the speech separation task that separates the speaker's identity information and content from the coupled speech signals. While in EEG systems, suppose models can be developed that separate the identity information from the task-related information. Then, the obtained identity information features can be further developed into an individualized and robust identification tool. In turn, the features of the new task-related information can be further developed into a cross-subject robust EEG classifier.

% \subsubsection{Low-shot Learning}
% In the future, data lightweight will be a significant trend in artificial intelligence, which also applies to EEG systems. Therefore, low shot learning techniques that aim to classify or recognize patterns in data with very few examples or "shots" are bound to be one of the techniques used to build robust EEG systems.

%  Low-shot learning in EEG systems is still challenging but holds great promise. Some new emerging low-shot techniques, such as contrastive learning \cite{shen2022contrastive} and meta learning \cite{li2021model}, have been just utilized in EEG systems. Till now, research on applying few shot learning in EEG systems is still in its infancy, and in the future, it has great potential to create robust EEG systems based on limited data.

\subsection{Building Interpretable and Robust EEG Systems}

% Building human-trusted EEG systems have been a long-term goal pursued by academics for many years. Using interpretability to identify potential problems and vulnerabilities in models can improve systems' robustness. In addition, with the addition of prior human knowledge and hidden semantics interpretation, the EEG systems can better learn from experts while the consumers can better understand the systems' mechanisms. This will be helpful for future academics to improve models and develop better EEG systems to meet consumers' requirements.
Building human-trusted EEG systems have been a long-term goal pursued by academics for many years. Using interpretability to identify potential problems and vulnerabilities in models can improve the robustness of EEG systems. In addition, incorporating prior human knowledge and interpreting hidden semantics can allow systems to better learn from experts. It also enables consumers to better understand how the systems work. This will be helpful for future academics to improve models and develop better EEG systems to meet consumers' requirements.

\section{Conclusion}

The interpretability and robustness of AI models in electroencephalogram (EEG) systems is growing in importance and urgency. They ensure the trustworthiness and reliability of EEG systems, and greatly contribute to understanding the models and reproducing the results. This survey pioneers a comprehensive overview of the interpretable and robust AI techniques designed explicitly for EEG systems. We provide a systemic perspective of this critical field, summarize a wide range of available techniques and tools, and offer an authoritative reference for researchers and practitioners. We introduce new and innovative taxonomies for interpretability and robustness in EEG systems. Throughout the survey, we summarize the most representative works based on their distinctive contributions, inventive mechanisms, or potential influence on the development of EEG systems. We analyze the technical details, properties and limitations of different works within each category, and also compare their differences across different categories. Highlighting these emerging techniques offers insight into the latest trends of this research area, and also provides a glimpse into their enormous potential and implications for the broader field. Furthermore, we discuss some unsolved problems and promising future directions in EEG systems. By identifying these issues and potential solutions, we hope to ignite discussions and inspire further research, ultimately pushing the boundaries of what EEG systems can achieve. In conclusion, this survey serves as an exhaustive guide for understanding, evaluating and advancing the realm of interpretability and robustness in EEG systems.

% or


% \appendices

% \section{}

% \section*{Acknowledgment}


\ifCLASSOPTIONcaptionsoff
  \newpage
\fi

\bibliographystyle{IEEEtran}
\bibliography{tnnls}

% \end{IEEEbiography}


%\newpage



%\vfill


\end{document}